\documentclass[
 aip,
 amsmath,amssymb,
 reprint,
]{revtex4-1}

\usepackage{graphicx}% Include figure files
\usepackage{dcolumn}% Align table columns on decimal point
\usepackage{bm}% bold math
\usepackage{cancel}
\usepackage[utf8]{inputenc}
\usepackage[T1]{fontenc}
\usepackage{mathptmx}
\usepackage{etoolbox}
\usepackage[dvipsnames,table,xcdraw]{xcolor}
\usepackage{soul}

\DeclareMathOperator*{\argmax}{arg\,max}
\DeclareMathOperator{\sech}{sech}
\newcommand{\vect}[1]{\boldsymbol{#1}}
\newcommand{\diff}{\mathrm{d}}

\makeatletter
\def\@email#1#2{%
 \endgroup
 \patchcmd{\titleblock@produce}
  {\frontmatter@RRAPformat}
  {\frontmatter@RRAPformat{\produce@RRAP{*#1\href{mailto:#2}{#2}}}\frontmatter@RRAPformat}
  {}{}
}%
\makeatother
\begin{document}

\preprint{AIP/123-QED}

\title[]{Frequency and field-dependent response of confined electrolytes from Brownian dynamics simulations}

\author{Th\^e Hoang Ngoc Minh}
\affiliation{ 
Sorbonne Universit\'e, CNRS, Physicochimie des \'Electrolytes et Nanosyst\`emes Interfaciaux, F-75005 Paris, France
}

\author{Gabriel Stoltz}%
\affiliation{CERMICS, Ecole des Ponts, Marne-la-Vallée, France}
\affiliation{MATHERIALS team-project, Inria Paris, France}

\author{Benjamin Rotenberg}%
\affiliation{ 
Sorbonne Universit\'e, CNRS, Physicochimie des \'Electrolytes et Nanosyst\`emes Interfaciaux, F-75005 Paris, France
}
\affiliation{R\'eseau sur le Stockage Electrochimique de l'Energie (RS2E), FR CNRS 3459, 80039 Amiens Cedex, France}

\date{\today}

\begin{abstract}
Using Brownian dynamics simulations, we investigate the effects of confinement, adsorption on surfaces and ion-ion interactions on the response of confined electrolyte solutions to oscillating electric fields in the direction perpendicular to the confining walls. Nonequilibrium simulations allow to characterize the transitions between linear and nonlinear regimes when varying the magnitude and frequency of the applied field but the linear response, characterized by the frequency-dependent conductivity, is more efficiently predicted from the equilibrium current fluctuations. To that end, we (rederive and) use the Green--Kubo relation appropriate for overdamped dynamics, which differs from the standard one for Newtonian or underdamped Langevin dynamics. This expression highlights the contributions of the underlying Brownian fluctuations and of the interactions of the particles between them and with external potentials. While already known in the literature, this relation has rarely been used to date, beyond the static limit to determine the effective diffusion coefficient or the DC conductivity. The frequency-dependent conductivity always decays from a bulk-like behavior at high frequency to a vanishing conductivity at low frequency due to the confinement of the charge carriers by the walls. We discuss the characteristic features of the crossover between the two regimes, most importantly how the crossover frequency depends on the confining distance and the salt concentration, and the fact that adsorption on the walls may lead to significant changes both at high- and low-frequencies. Conversely, our results illustrate the possibility to obtain information on diffusion between walls, charge relaxation and adsorption by analyzing the frequency-dependent conductivity.
\end{abstract}

\maketitle

\section{Introduction}

The dynamics of bulk and confined electrolytes play an essential role in fields as diverse as electrochemical energy storage in batteries and supercapacitors~\cite{simon_perspectives_2020}, energy harvesting from salinity gradients (so-called blue or osmotic energy) and desalination~\cite{elimenech_future_2011,siria_new_2017,simoncelli_blue_2018,xiao_ion_2019,liu_carbon_2021}, biological and bio-inspired systems~\cite{robin_modeling_2021}, or contaminant transport and retention (\textit{e.g.} by clay minerals) in the environment~\cite{liu_molecular-level_2022}. The motion of ions in colloidal suspensions, polyelectrolyte solutions or porous materials contributes significantly to the overall electric response of these complex systems~\cite{pride_governing_1994,bordi_dielectric_2004,grosse_dielectric_2010,zhou_dielectric_2012,zhou_dynamic_2013,zhou_computer_2013,merlin_electrodynamics_2014}, characterized by the complex, frequency-dependent conductivity or permittivity, or observables quantifying more involved properties such as electro-acoustic couplings. The interpretation of electrochemical impedance spectroscopy, routinely used to characterize energy storage devices, requires taking into account charge transport and interfacial processes occuring at the surface of electrodes~\cite{wang_electrochemical_2021,vivier_impedance_2022}. 

The equilibrium fluctuations of the ionic current, quantified \textit{e.g.} by its power spectral density, encode information on the underlying microscopic dynamics of the ions in the solvent. Following earlier studies of bulk charge transport~\cite{hooge_1_f_1970,vasilescu_electrical_1974}, the ``electrical noise'' measured through nanopores~\cite{hoogerheide_probing_surface_2009,siria_giant_2013,laszlo_decoding_2014,heerema_1_f_2015,secchi_scaling_2016} and in (nano)electrochemical devices ~\cite{bertocci_noise_1995,zevenbergen_electrochemical_2009,mathwig_electrical_2012} can in principle be used to infer information on charge transport and interfacial processes. However this step requires disentangling the contributions of the various processes that contribute to the current fluctuations, such as diffusion, migration, advection, adsorption/desorption on surfaces, or (redox) reactions, and resorting to modelling is necessary to interpret the experiments.

Since the pioneering work of Debye, H\"uckel and Onsager, the canonical theory of ion transport in bulk electrolytes relies on the description of ions experiencing the effects of an implicit solvent~\cite{onsager_report_1927,onsager_deviations_1934,onsager_relaxatio_1957}: thermal fluctuations (Brownian motion) and friction, screening of electrostatic interactions (permittivity), and hydrodynamic interactions (viscosity). Various levels of refinement can be adopted, for example to capture the short-range repulsion between ions due to their finite size~\cite{friedman_limiting_1964,friedman_relax_1965,bernard_self_diffusion_1992,bernard_conductance_1992,bernard_binding_msa_1996,dufreche_analytical_2005}. The more recent development of stochastic Density Functional Theory~\cite{kawasaki_stochastic_1994,dean_langevin_1996} also allowed to recover earlier results for electrolytes and provide in principle a way to couple consistently the ionic and solvent fluctuations~\cite{demery_conductivity_2016,peraud_fluctuating_2017,donev_fluctuating_hydro_2019,avni_conductivity_2022}. Additionally, Brownian descriptions have been used to predict the frequency-dependent conductivity or electro-acoustic couplings in bulk electrolytes~\cite{anderson_debye_falkenhagen_1994,durand-vidal_acoustophoresis_1995,chandra_frequency_2000,yamaguchi_theoretical_frequency_dependent_conductivity_2007}. 

For confined electrolytes, the above-mentioned experiments prompted theoretical efforts to model current fluctuations through nanopores~\cite{kowalczy_modeling_2011,zorkot_current_2016,zorkot_power_2016,zorkot_current_sPNP_2018,gravelle_adsorption_2019,mahdisoltani_long_2021,marbach_intrinsic_2021}. The charging dynamics in electrochemical devices and its link to charge or current fluctuations is an old problem, which regained interest in the context of nanocapacitors, with various approaches from molecular simulations to meso- and macroscopic scale studies~\cite{limmer_charge_2013,scalfi_charge_2020,scalfi_molecular_2021,PiredduArXiv,bazant_diffuse-charge_2004,janssen_transient_2018,chassagne_compensating_2016,asta_lattice_2019,ma_dynamic_2022}. The possibility to extract information on the microscopic dynamics of ions from the frequency-dependent permittivity of materials was examined \textit{e.g.} for salt-free charged lamellar systems (such as biological membranes, clay-like minerals), with an analytical solution of the coupled equations describing diffusion, electrostatics at the mean-field level, and adsorption/desorption of counterions~\cite{rotenberg_frequency-dependent_2005}.

As in other contexts, numerical simulations on various scales can be used to describe the dynamics in electrolytes beyond the regimes, usually limited to low concentrations, where analytical descriptions apply~\cite{pagonabarraga_recent_2010,rotenberg_coarse-grained_2010,site_multiscale_2012,rotenberg_electrokinetics_2013}. While some studies used molecular simulations to investigate the frequency-dependent conductivity of electrolyte solutions~\cite{chandra_frequency_1993,tang_non-equilibrium_2002}, the natural approach to capture the above-mentioned effects of an implicit solvent (even though the treatment of hydrodynamic interactions requires special care), is to describe the motion of ions by Brownian dynamics, \textit{i.e} overdamped Langevin dynamics~\cite{risken_fokker_1996}. It has been successfully used to investigate transport in bulk electrolyte solutions and suspensions of charged nanoparticles~\cite{jardat_transport_1999,jardat_brownian_2000,jardat_brownian_2004,dahirel_two_scale_2009,yamaguchi_brownian_2011}, or confined electrolytes~\cite{jardat_self-diffusion_2012}. Various nonlinear responses have been identified depending in particular on the importance of ion-ion interactions (weak \textit{vs} strong electrostatic coupling regime, hydrodynamic interactions), the magnitude of the applied external field and confinement~\cite{netz_conduction_2003,lobaskin_diffusive-convective_2016}.

Here, we use Brownian dynamics simulations to investigate the field- and frequency-dependent response of confined electrolytes, and assess the effects of the geometry, adsorption on the confining walls and ion-ion interactions. We show the benefits of using linear response theory to predict the frequency-dependent conductivity from equilibrium simulations, using the Green--Kubo expression appropriate for overdamped Langevin dynamics~\cite{felderhof_linear_1983,felderhof_linear_1987,contreras_aburto_unifying_2013}. Even though the static limit of the latter has already been used in Brownian dynamics simulations to determine the effective diffusion coefficient or the static conductivity~\cite{jardat_transport_1999,jardat_brownian_2000,jardat_brownian_2004,dahirel_two_scale_2009}, the frequency-dependent result (for which we provide a slightly different derivation inspired from Ref.~\citenum{joubaud_langevin_2015}) does not seem to have been much exploited in the literature. Section~\ref{sec:BD} introduces the model of confined electrolyte solutions, the relevant observables to characterize the field- and frequency-dependent response to an applied electric field and the simulation details. All results are then presented in Section~\ref{sec:Results}.

%%%%%%%%%%%%%%%%%%%%%%%%%%%%%%%%%%%%%%%%%%%%%%%%%%%%%%%%%%%%%%%%%%%%%%%%%%%%%%%%%%%%
\section{Brownian dynamics of confined electrolyte solutions}
\label{sec:BD}

In order to investigate the influence of confinement, adsorption on surfaces and ion-ion interactions on the response of electrolyte solutions to an oscillatory electric field, we consider Brownian particles confined between two parallel walls and compare several models introducing progressively the above physical features, as described in Section~\ref{sec:BD:Model}. We then introduce the relevant physical observables in Section~\ref{sec:BD:Conductivity} and simulations details in Section~\ref{sec:BD:SimuationDetails}.

%%%%%%%%%%%%%%%%%%%%%%%%%%%%%%%%%%%%%%%%%%
\subsection{Model}
\label{sec:BD:Model}

\begin{figure}[ht!]
    \begin{center}
        \includegraphics[width = 0.4\textwidth]{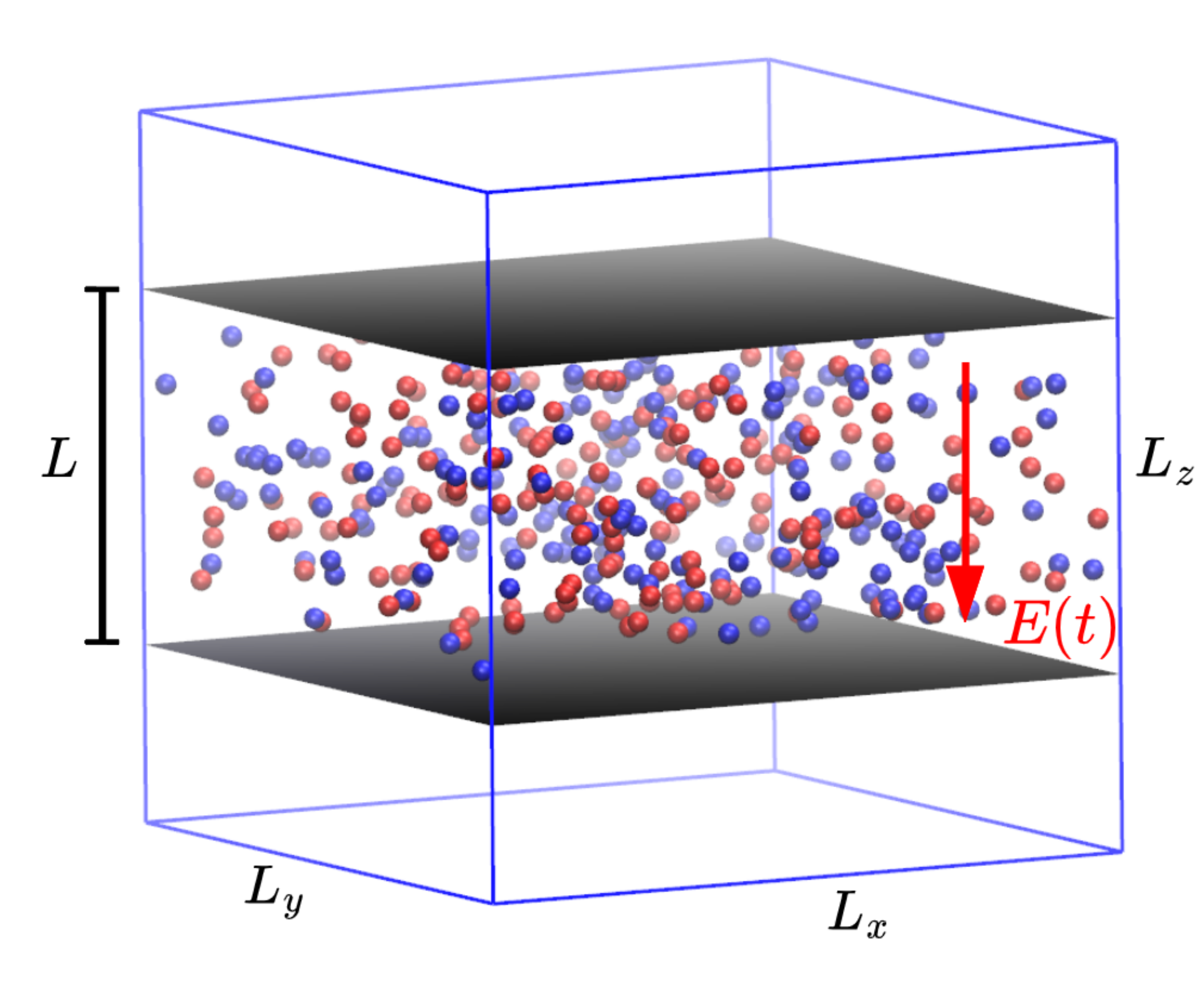}
    \end{center}
    \caption{The simulated systems consist of a 1:1 electrolyte, described by ions in an implicit solvent, confined between parallel walls separated by a distance $L$. The box is cubic with dimensions $L_x=L_y=L_z=L_\mathrm{box}$ and periodic boundary conditions are used in the directions along the planes. 3D periodic boundary conditions with a slab correction are used to compute electrostatic interactions, when present, to model a system with periodicity in the $\hat{x}$ and $\hat{y}$ directions only (note that the present study is limited to the simple case where the permittivity of the confining medium is equal to that of the confined fluid, see text). Ions interact with the confining walls via a short-range external potential which depends only on $z$, is repulsive in general and may also include an attractive part to model adsorption (see Eq.~\eqref{eq:Walls:Vads}). We investigate the response of the system to an oscillatory electric field $E(t)=E_0 \sin\omega t$ in the direction perpendicular to the walls.}
    \label{fig:snapshot}
\end{figure}

We consider an aqueous NaCl electrolyte solution confined in a slit pore between walls separated by a distance $L$, as illustrated in Figure~\ref{fig:snapshot}. The simulated systems consist of $N$ ion pairs with ionic charges $q_{i} = Z_i e $ (with valencies $Z_i=+1$ for cations and -1 for anions, respectively, and $e$ the elementary charge), in an implicit solvent characterized by its relative permittivity $\varepsilon_r=78.5$. The ions are described as Brownian particles with a diffusion coefficient $D_i$. Their positions $\vect{r}_i$ evolve according to the overdamped Langevin equation: 
\begin{equation}
    \dot{\vect{r}}_i =  \beta D_i \vect{F}_i + \sqrt{2 D_i}\hat{\vect{\eta}}_i
    \label{eq:BD}
\end{equation}
where $\beta=1/k_{\rm B}T$, with $k_{\rm B}$ Boltzmann's constant and $T$ the temperature, $\hat{\vect{\eta}}_i$ is a Gaussian white noise, and the force $\vect{F}_i$ acting on the ions may include the interactions between them and with the walls, as well as the force $q_i \vect{E}$ due to the applied time-dependent electric field in the $z$ direction with magnitude
\begin{equation}
    E(t)= H(t)E_0\sin\omega t
    \, ,
    \label{eq:Eoft}
\end{equation}
where the Heaviside function $H$ indicates that the field is applied only after some initial time taken as $t=0$ (the system is at equilibrium without field for $t<0$). For the model system considered here (whose limitations are emphasized below), the response to an electric field parallel to the walls does not significantly differ from that of a bulk electrolyte and will not be discussed here.

When interactions between ions are considered (as discussed below, we also study as a reference the case of ideal, \textit{i.e.} non-interacting, particles), they interact via pairwise additive potentials including electrostatic interactions, screened by the solvent, and a Week--Chandler--Andersen (WCA) potential to describe short-range repulsion:
\begin{equation}
v_{ij}(r) = \displaystyle \frac{q_iq_j}{4\pi\varepsilon_0\varepsilon_r r}
+ v_{ij}^{\mathrm{WCA}}(r),
\label{eq:PairPotential}
\end{equation}
with $\varepsilon_0$ the vacuum permittivity and
\begin{equation}
v_{ij}^{\mathrm{WCA}}(r) =
    \left\lbrace
    \begin{matrix}
    v_{ij}^\mathrm{LJ}(r) \, - \, v_{ij}^\mathrm{LJ}(r^*) \ &, \ r \leq r^*, \\
    0 \ &, \ r > r^*,
    \end{matrix}
    \right.
\end{equation}
with the Lennard--Jones (LJ) potential
\begin{equation}
v_{ij}^\mathrm{LJ}(r) =
4\epsilon_{ij} \left[ \left(\frac{\sigma_{ij}}{r}\right)^{12} -  \left(\frac{\sigma_{ij}}{r}\right)^6 \right],
\end{equation}
and $r^*=2^{1/6}\sigma_{ij}$ the position of the minimum of $v_{ij}^\mathrm{LJ}$. The LJ energy and diameter $\epsilon_{ij}$ and $\sigma_{ij}$ are computed from the corresponding parameters for ions $i$ and $j$ using the Lorentz--Berthelot mixing rules. 

\begin{figure}[ht!]
    \begin{center}
        \includegraphics[width = 0.4\textwidth]{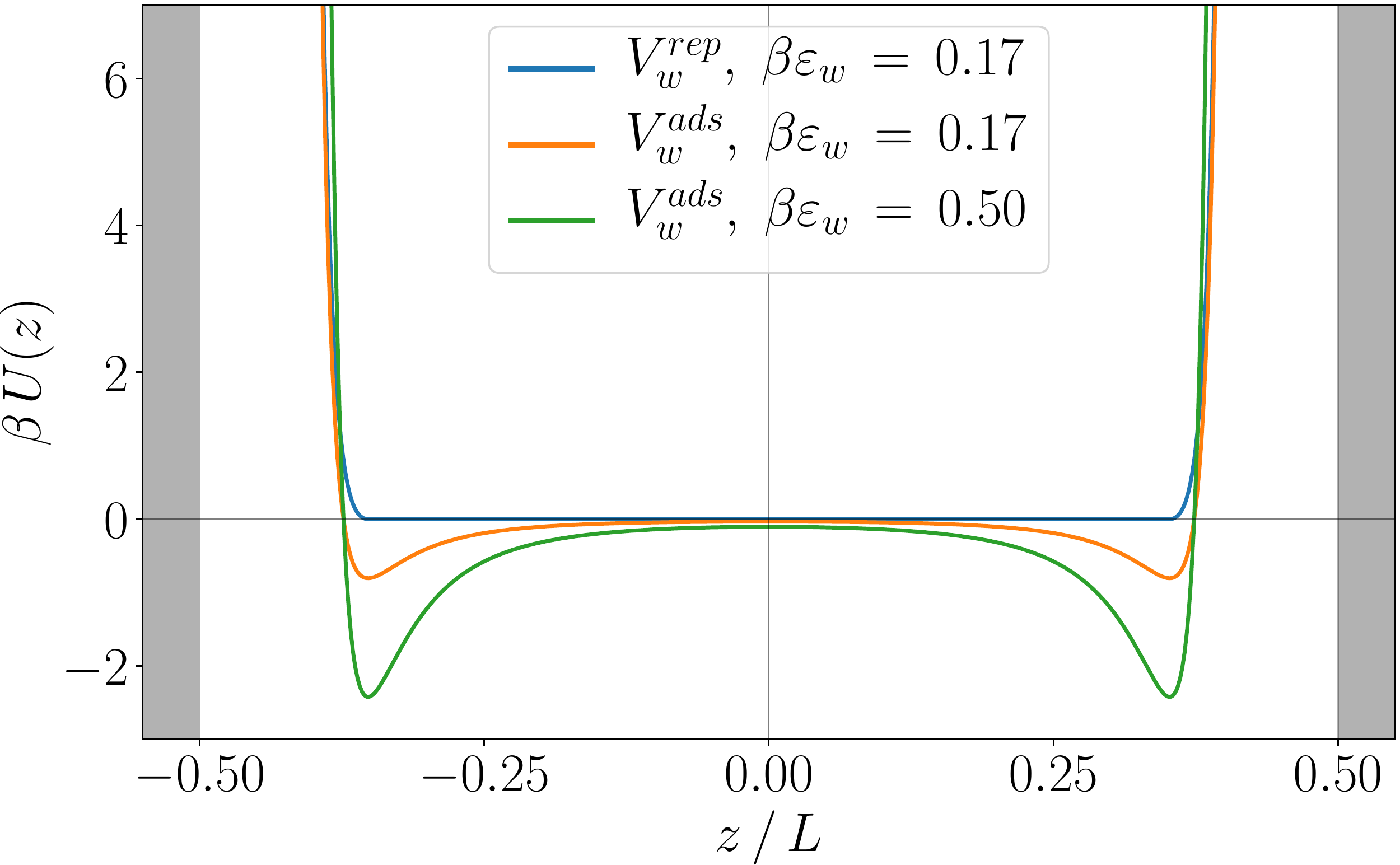}
    \end{center}
    \caption{
    Illustration of the ion-wall potential for $\sigma_\mathrm{w} / L=0.3$, with $L$ the distance between the two walls and $\sigma_\mathrm{w}$ the characteristic length entering in the potential. The orange and blue lines show the full ($V_\mathrm{w}^\mathrm{ads}$, Eq.~\eqref{eq:Walls:Vads}) and truncated-shifted ($V_\mathrm{w}^\mathrm{rep}$, Eq.~\eqref{eq:Walls:Vrep}) potential for $\beta \epsilon_\mathrm{w}=0.17$, while the green line corresponds to $V_\mathrm{w}^\mathrm{ads}$ for $\beta \epsilon_\mathrm{w}=0.50$.
    }
    \label{fig:potentials}
\end{figure}

Each ion interacts with the two walls placed at $z=\pm L/2$ through the potential
\begin{equation}
    U(z) \ = \ V_\mathrm{w}( z + L/2) \, + \, V_\mathrm{w}( z - L/2),
    \label{eq:Walls}
\end{equation}
which depends only on $z$. The chosen form arises from the integration of LJ interactions with atoms in a closed-packed face centered cubic lattice, with lattice parameter $\sigma_\mathrm{w}$, cut along a (100) face. The resulting potential includes short-range repulsion and an attractive well leading to adsorption~\cite{steele_physical_interaction_of_gases_1973,steele_interaction_of_rare_gas_1978,magda_molecular_1985,magda_erratum_1986}.
\begin{align}
V_\mathrm{w}^\mathrm{ads}(z) &=
2 \, \pi \, \epsilon_\mathrm{w} \displaystyle
\left[
\frac{2}{5} \left( \frac{\sigma_\mathrm{w}}{z}\right)^{10} \,
- \left( \frac{\sigma_\mathrm{w}}{z}\right)^{4}
\right. \nonumber \\ 
 & \left. \qquad\qquad\qquad \displaystyle
 - \frac{\sqrt{2}}{3 \left(  z/ \sigma_\mathrm{w} + 0.61/\sqrt{2}  \right)^3}
\right]
\ ,
\label{eq:Walls:Vads}  
\end{align}
where the energy $\epsilon_\mathrm{w}$ tunes the strength of the ion-wall attraction.
In order to model purely repulsive walls, we use the same truncation and shifting as for the short-range ion-ion interactions, \textit{i.e.}
\begin{equation}
    V_{w}^\mathrm{rep}(z)  \ = \ 
    \left\lbrace
    \begin{matrix}
    V_\mathrm{w}^\mathrm{ads}(z) \, - \, V_\mathrm{w}^\mathrm{ads}(z^*) \ &, \ z \leq z^*, \\
    0 \ &, \ z > z^*,
    \end{matrix}
    \right.
\label{eq:Walls:Vrep} 
\end{equation}
with $z^*\approx 0.987\sigma_\mathrm{w}$ corresponding to the minimum of $V_\mathrm{w}^\mathrm{ads}$. Fig.~\ref{fig:potentials} illustrates these ion-wall potentials for some cases considered (described in more detail in Section~\ref{sec:BD:SimuationDetails}).

Despite its limitations (see \textit{e.g.} Ref.~\citenum{banerjee_ions_2019} for a recent discussion), the overdamped Langevin equation~\eqref{eq:BD} provides a good description of the dynamics of monoatomic ions for the prediction of the static conductivity or the effective diffusion coefficient of ions~\cite{jardat_transport_1999,jardat_brownian_2004,dahirel_two_scale_2009}. Since we focus here on the frequency-dependent response, we should emphasize that it neglects all inertial effects, corresponding to the relaxation of the ion velocity. This occurs on a time scale $\tau_i=\beta m_i D_i$, with $m_i$ the mass of the ion. For frequencies such that $\omega\tau_i\gtrapprox1$, Eq.~\eqref{eq:BD} should be replaced by an underdamped Langevin dynamics or even a molecular description. However for NaCl, $\tau_i$ is in the range of 10-25~fs, corresponding to frequencies $f=\omega/2\pi$ as high as 1~THz, which will not be considered here. The present model also neglects the frequency dependence of the solvent permittivity, which decreases significantly for frequencies larger than 10~GHz.
 
Another important aspect neglected in the present Brownian description is the role of hydrodynamics, which modifies the dynamics of ions in several ways. In the bulk, hydrodynamic interactions between ions impact (even more so for increasing salt concentration) their self-diffusion as well as the conductivity of the solution via the so-called electrophoretic effect (see \emph{e.g.} Ref.~\citenum{dufreche_analytical_2005}). A second point related to the coupling of the ions with the solvent is that the presence of solid walls perturbs the hydrodynamic flows, an effect which is captured at the continuum level by imposing a vanishing normal velocity (no flux into the solid) and either no-slip or slip boundary condition for the tangential velocity. This solvent-mediated solute-wall coupling changes the mobility (or the diffusion coefficient) of the solute, which becomes anisotropic, to an extent which depends on the distance to the wall (or the relative position with respect to both walls under confinement) and on the hydrodynamic boundary condition~\cite{saugey_diffusion_2005, swan_simulation_2007, delong_brownian_dynamics_2015}. We further note that the presence of interfaces leads to the layering of the solvent, which may also impact the dynamics of ions in the close vicinity of the surface beyond the continuum hydrodynamic description. The decrease of the collective or individual mobility as solutes approach a surface has also been reported in molecular simulations (see \emph{e.g.} Refs.~\citenum{marry_structure_2008, simonnin_mineral_2018, mangaud_sampling_mobility_2020}), and hydrodynamics also explains finite-size effects due to the periodic boundary conditions in the directions parallel to the walls on the diffusion coefficient of confined particles~\cite{simonnin_diffusion_2017}.

More generally, the couplings between ions and solvent give rise to so-called electrokinetic effects such as electro-osmosis (solvent flow induced by an electric field), which arise at charged interfaces. Such couplings may develop over various scales and pose specific challenges for modelling (see \emph{e.g.} Refs.~\citenum{pagonabarraga_recent_2010, bonthuis_unraveling_2012, rotenberg_electrokinetics_2013, cui_gle_2018, ladiges_discrete_ion_2021, ladiges_modeling_electrokinetic_2022, tischler_thermalized_2022}). In order to address progressively the contributions of confinement, adsorption on the walls and ion-ion interactions on the frequency-dependent conductivity, we leave the above effects related to the hydrodynamic coupling to the solvent for future work. Such an approximation is of course expected to be less accurate as the salt concentration increases and the distance between walls decreases. We note that for sufficiently low applied fields normal to the walls (in particular in the linear response regime), there is no flow in that direction (see \emph{e.g.} Ref.~\citenum{asta_lattice_2019}), so that neglecting hydrodynamics might be sufficient, but that convective flows may develop for large fields~\cite{lobaskin_diffusive-convective_2016}.

Finally, an interface between media with different dielectric properties modifies electrostatic interactions, which can be accounted for by the concept of image charges, and in practice require special care when performing simulations of confined electrolytes. The two extreme cases are that of a liquid/air interface, where the relative permittivity $\epsilon_r$ of the liquid is larger than that of air (close to that of vacuum), especially for water, and that of an interface between a liquid and a metallic wall (which can be considered as the limit of infinite permittivity). It is for example well known that such dielectric jumps at water/air, water/oil or water/membrane interfaces play an important role on the distribution of ions and on the interfacial tension, even when the interface is charge-neutral~\cite{croxton_ionic_1981, levin_surface_2001, bier_liquid-liquid_2008, rotenberg_coarse-grained_2010, buyukdagli_variational_2010, levin_ions_2014}. They can also influence the ionic mobility near walls in non-trivial ways~\cite{antila_dielectric_2018} (see also the above discussion of hydrodynamics). Efficient algorithms have been introduced to deal with such dielectric jumps in particle-based simulations~\cite{tyagi_electrostatic_layer_2008, arnold_efficient_2013, barros_efficient_2014, liang_harmonic_2020, yuan_particleparticle_2021, maxian_fast_spectral_2021, liang_hsma_2022}. For simplicity, we will only consider the special case with no dielectric jump at the interface, \emph{i.e.} when the permittivity of the surrounding medium is equal to that of the confined fluid. While this is admittedly not the most relevant case for aqueous electrolytes, this can be considered as a first step to describe the effect of electrostatic interactions between ions and we leave the effect of the influence of the dielectric jump for further study.

With all these caveats in mind, the present model provides a good starting point for a systematic analysis of the field- and frequency-dependent response of confined electrolytes to an oscillatory electric field.

%%%%%%%%%%%%%%%%%%%%%%%%%%%%%%%%%%%%%%%%%%
\subsection{Field- and frequency-dependent conductivity}
\label{sec:BD:Conductivity}

The conductivity quantifies the electric current induced by an applied electric field. The instantaneous electric current is generally determined from the velocity of charged particles as $\sum_i q_i \vect{v}_i(t)$, where the sum runs over ions $i$. However, the velocity is ill-defined for the overdamped Langevin dynamics (Eq.~\eqref{eq:BD}) and one considers instead the hydrodynamic velocity $\beta D_i \vect{F}_i(t)$, reached by the ions within a time scale $\tau_i=\beta m_i D_i$, to define the electric current
\begin{equation}
    J_\mathrm{el}(t) = \sum_i \beta D_i q_i F_{i,z}(t)
    \; ,
    \label{eq:JelBD}
\end{equation}
where the force in the $\hat{z}$ direction includes the interactions with the other ions and the walls, as well as the force $q_i E(t)$ arising from the applied field. For a bulk ideal electrolyte (\textit{i.e.} unconfined non-interacting ions), the latter is the only contribution to the current and $J_\mathrm{el}^{\mathrm{NE}}(t)=V\sigma^{NE} E(t)$, with $V$ the volume of the system and the Nernst-Einstein conductivity
\begin{equation}
\sigma^{\mathrm{NE}} = \frac{\beta}{V}\sum_i D_i q_i^2 
= \frac{\beta e^2 N \left( D_+ + D_- \right) }{V} 
\; ,
   \label{eq:NE}
\end{equation}
the second equality holding only for the present case of a 1:1 electrolyte, with diffusion coefficients $D_+$ and $D_-$ for cations and anions, respectively.

Starting from an equilibrium configuration in the absence of the field, applying from $t=0$ an external field $E(t)= E_0\sin\omega t=\operatorname{Im}[E_0 e^{\mathrm{i}\omega t}]$, leads after a transient regime to a stationary regime where the current is also periodic with the same period $T=2\pi/\omega$. In the limit of vanishingly small perturbation, $E_0\to 0$, the response is linear so that the current oscillates at the same frequency. The complex, frequency-dependent conductivity $\tilde{\sigma}(\omega)$ can then be defined from the stationary current by
\begin{equation}
  \lim_{E_0 \to 0} \frac{J_\mathrm{el}(t)}{ V E_0} = \operatorname{Im}\left[ \tilde{\sigma}(\omega)  e^{\mathrm{i}\omega t} \right]
  \; .
  \label{eq:LinearResponse}
\end{equation}
The conductivity is in general a complex quantity, but its real and imaginary parts are linked by the Kramers-Kronig relations. In practice, it is therefore sufficient to study the real part $\sigma(\omega)=\operatorname{Re}[\tilde{\sigma}(\omega)]$ to fully characterize the frequency-dependence.

While it is possible to determine the conductivity in nonequilibrium simulations in the presence of an applied field, for reasons discussed below it is however standard practice in molecular simulations to determine the linear response from the equilibrium fluctuations, \textit{i.e.} in the absence, using Green--Kubo (GK) relations. In molecular dynamics (MD) simulations, the complex conductivity is obtained from the Laplace transform of the current autocorrelation function (ACF) as~\cite{hansen_mcdonald_theory_of_simple_liquids_book}
\begin{equation}
    \tilde{\sigma}_\mathrm{MD}^\mathrm{GK}(\omega) =
    \frac{\beta}{V} \int_0^{\infty} \left\langle J_\mathrm{el}^\mathrm{MD}(0) J_\mathrm{el}^\mathrm{MD}(t) \right\rangle_{0} \mathrm{e}^{-\mathrm{i}\omega t}\diff t
    \label{eq:GK_conductivity_MD}
\end{equation}
with $J_\mathrm{el}^\mathrm{MD}(t) = \sum_i q_i v_{i,z}(t)$ (for a bulk isotropic system, it is often written as an average over the 3 directions of space). However, in Brownian dynamics the relevant current is given instead by Eq.~\eqref{eq:JelBD} and linear response theory leads to a different expression of the complex conductivity:
\begin{equation}
     \tilde{\sigma}_\mathrm{BD}^\mathrm{GK}(\omega) =
     \sigma^{\mathrm{NE}} 
     - \dfrac{\beta}{V} \int_0^{\infty}
     \left\langle J_\mathrm{el}^0(0) \,J_\mathrm{el}^0(t) \right\rangle_0  \, \mathrm{e}^{-\mathrm{i} \omega t} \, \diff t
     \; .
    \label{eq:GK_conductivity_BD}
\end{equation}
Note that the current ACF is at equilibrium, \textit{i.e.} in the absence of applied field, so that the only contributions to the electric current in Eq.~\eqref{eq:JelBD} come from the interactions between the ions and with the walls (hence the notation~$J_\mathrm{el}^0$ to emphasize this point). Eq.~\eqref{eq:GK_conductivity_BD} has been derived (in a simpler form for the sedimentation of a one-component system, but the extension is straightforward) by Felderhof and Jones~\cite{felderhof_linear_1983}. In Appendix~\ref{sec:Appendix:GK} we provide a slightly different derivation based on the Fokker--Planck (FP) equation for the evolution of the probability distribution $\rho(\vect{R},t)$, with $\vect{R}=\left\{ \vect{r}_i \right\}_{i=1\dots 2N}$ corresponding to the trajectories of ions following Eq.~\eqref{eq:BD}. The FP equation can be solved directly in the case where the ions interact only with the walls but not with each other. This provides an alternative route to compute the frequency-dependent linear response given by Eq.~\eqref{eq:LinearResponse} to validate the numerical method in this ideal case. Details on this approach are provided in Appendix~\ref{sec:Appendix:FP}.

Even though the static limit of Eq.~\eqref{eq:GK_conductivity_BD} (\textit{i.e.} the limit $\omega\to 0$) has already been used in Brownian dynamics simulations to determine the effective diffusion coefficient or the static conductivity~\cite{jardat_brownian_2004,jardat_transport_1999,dahirel_two_scale_2009}, the frequency-dependent result does not seem to have been much exploited in the literature. This expression highlights the fact that the frequency-dependence arises only from the interactions. It follows immediately that for case of noninteracting particles and smooth walls, there is no force in the directions parallel to the walls and the corresponding conductivity (which will not be further discussed here) reduces to $\sigma^{\mathrm{NE}}$ for all frequencies. Anticipating the results, in contrast the confinement of the particles by the walls results in all the considered cases in $\lim_{\omega\to0}\sigma(\omega)=0$ in the $\hat{z}$-direction. Such a separation between the underlying Brownian motion and the interactions has also been introduced to determine the velocity ACF of Brownian hard spheres (from the time-dependent mean-square displacement) in Ref.~\citenum{mandal_persistent_2019}; however the possible connection with the above result was not noted and the assumption of neglecting a correlation between the ideal and interaction term (which was verified numerically) was necessary.

The nonlinear response is \emph{a priori} multimodal and can be analyzed using various properties of the stationary regime of $J_\mathrm{el}(t)$ over a period $T$. In order to facilitate the comparison with the linear regime, we use the following linear combination of Fourier coefficients of the current at the frequency of the applied field 
\begin{equation}
    J_\mathrm{el}(\omega)= \displaystyle\frac{2}{T}\int_0^T J_\mathrm{el}(t)\sin\omega t \,{\rm d}t 
    \label{eq:Jomega}
\end{equation}
to define the field- and frequency-dependent conductivity as:
\begin{equation}
    \sigma(E_0,\omega) \equiv \displaystyle \frac{J_\mathrm{el}(\omega)}{ V E_0}
    \, .
    \label{eq:SigmaEandOmega}
\end{equation}
In the limit $E_0\to0$, one recovers $\sigma(E_0,\omega)\to\sigma(\omega)$ by definition (see Eq.~\eqref{eq:LinearResponse}, which states that $J_\mathrm{el}(t)$ is, in the small field limit, the sum of~$\sigma(\omega) \sin(\omega t)$ and a term proportional to $\cos(\omega t)$, which disappears when integrating in~\eqref{eq:Jomega}). Other properties to characterize the nonlinear response are defined and discussed in Appendix~\ref{sec:Appendix:NL}.

%%%%%%%%%%%%%%%%%%%%%%%%%%%%%%%%%%%%%%%%%%
\subsection{Simulation details}
\label{sec:BD:SimuationDetails}

\begin{table}[ht!]
    \centering
  \begin{tabular}{|c|c|c|c|c|}
    \hline
     System & Ion pairs $N$ & $L_\mathrm{box}$ (\AA) & $L$ (\AA) \\
    \hline
    \hline
     Rep. walls & $200$ & $40$ & $\left\lbrace 20, \, 16,\,12 \right\rbrace$ \\
    \hline
     Attr. walls & $200$ & $40$ & $20$ \\
    \hline
    \hline
     Electrolyte I & $\left\lbrace 30, \, 60, \, 120, \, 240, \, 480 \right\rbrace$ & $100$ & $100$\\
    \hline
     Electrolyte II & $\left\lbrace 20, \, 45, \, 90, \, 180, \, 360 \right\rbrace$ & $200$ & $35$\\
    \hline
     Electrolyte III & $\left\lbrace 45, \, 90, \, 180 \right\rbrace$ & $800$ & $35$\\
    \hline
  \end{tabular}
    \caption{Simulated systems. 
    In the first two cases, the ions are treated as ideal particles and their interactions with the walls can be repulsive (Eq.~\eqref{eq:Walls:Vrep}) or attractive (Eq.~\eqref{eq:Walls:Vads}). In the last three cases ("Electrolytes"), ions interact with each other via WCA and electrostatic interactions (see Eq.~\eqref{eq:PairPotential}) and repulsively with the walls. $N$ cations and $N$ anions are placed between the two walls, separated by a distance $L$ in a cubic box ($L_x=L_y=L_z=L_\mathrm{box}$), as shown in Fig.~\ref{fig:snapshot}. 
    }
    \label{tab:systems}
\end{table}

Table~\ref{tab:systems} summarizes the considered systems (see Fig.~\ref{fig:snapshot}), in terms of composition, dimensions and interactions. A first series allows to consider the effect of the confining distance $L$ between repulsive walls for ideal particles. The introduction of attractive walls then allows to consider, for a fixed $L$, the effect of adsorption. Finally, we investigate the effect of ion-ion interactions as a function of concentration $C_s$ for various $L$, with repulsive walls. The salt concentrations corresponding to the chosen number of ion pairs $N$, box size $L_\mathrm{box}$ and distance $L$ between walls range from $0.004$ to $0.8$~mol~L$^{-1}$. In all cases, periodic boundary conditions are applied in the directions $\hat{x}$ and $\hat{y}$ along the walls. For nonequilibrium simulations, an oscillatory electric field $E(t)=E_0 \sin\omega t$ is applied in the $\hat{z}$ direction perpendicular to the walls.

We consider diffusion coefficients $D_+=1.28~10^{-9}$~m$^2$.s$^{-1}$ for Na$^+$ and $D_-=1.77~10^{-9}$~m$^2$.s$^{-1}$ for Cl$^-$. These values were obtained for ions at infinite dilution and room temperature by MD simulations with an explicit solvent (SPC/E water model)~\cite{koneshan_solvent_1998}. The overdamped Langevin equation~\eqref{eq:BD} is solved numerically using the LAMMPS simulation package~\cite{thompson_LAMMPS_2022}, adapted with the overdampled BAOAB integrator~\cite{Leimkuhler_BAOAB_2013}. The interactions of the ions with the walls are given by Eqs.~\eqref{eq:Walls}, \eqref{eq:Walls:Vads} and~\eqref{eq:Walls:Vrep} with $\sigma_\mathrm{w}=3$~\AA, and $\epsilon_\mathrm{w}=0.1$ or $0.3$~kcal.mol$^{-1}$, corresponding to $\beta \epsilon_\mathrm{w}\approx 0.17$ or 0.5 at $T=300$~K (see Fig.~\ref{fig:potentials}). For the WCA interactions between ions, we use $\sigma_\mathrm{Na}=3$~\AA\ (resp. $\sigma_\mathrm{Cl}=3$~\AA) and $\epsilon_\mathrm{Na}=0.1$~kcal.mol$^{-1}$ (resp. $\epsilon_\mathrm{Cl}=0.1$~kcal.mol$^{-1}$), with the Lorentz-Berthelot mixing rule to compute the interactions between Na$^+$ and Cl$^-$. Long-range electrostatic interactions are computed with the P3M algorithm, using a cutoff of 15~\AA\ for the real-space part and a slab correction (with an equivalent system width of $3L_z$ including the empty slab) to model a system with periodicity in the $\hat{x}$ and $\hat{y}$ directions only~\cite{yeh_ewald_1999}. Note that this treatment of electrostatic interactions only applies to the special case where there is no dielectric jump between the confined liquid and the surrounding medium (see Section~\ref{sec:BD:Model}). 

For equilibrium simulations, we use a time step $\delta t=50$~fs, except for the electrolyte with the largest salt concentration ($C_s=0.8$~M) for which we use $\delta t=25$~fs. The total length of the trajectory is $T_\mathrm{tot}=50$~$\mu$s, except for the same system ($T_\mathrm{tot}=25$~$\mu$s) and in the case of ideal particles with the most attractive walls ($T_\mathrm{tot}=100$~$\mu$s). For each simulation, the trajectory is divided into $N_\mathrm{blocks}=100$ blocks (200 for the most attractive walls) on which the properties such as time-correlation functions or their Fourier transforms are computed. The reported results and uncertainties correspond to the average and the 95\% confidence interval computed using the standard deviation among blocks, respectively. The length of each block (500~ns for most systems, 250~ns for ideal particles with the most attractive walls) is much longer than all the correlation times in the various systems, so that the blocks can be considered as statistically independent from each other. For each block, the real part of the frequency-dependent conductivity is computed using a non-uniform sampling of the current ACF and numerical integration of Eq.~\eqref{eq:GK_conductivity_BD} using the trapezoidal rule for the considered frequencies. Specifically, the short-time behavior of the ACF (for $t<10$~ps) is estimated from the first 2000 steps of each block with the current evaluated at every step, while the rest of the ACF is estimated from samples of the current every 100 time steps over the whole block. 

Nonequilibrium simulations are performed with repulsive walls separated by $L=20$ \AA~ with 200 pairs of particles (first entry of Table~\ref{tab:systems}) and in the presence of a field $E(t) = E_0\sin(\omega t)$, using a time step $\delta t=50$~fs, with a total trajectory length $T_{tot}=1$~$\mu$s. The results presented in Fig.~\ref{fig:NEQ_currents} are obtained for field magnitudes $E_0 \in \left\lbrace 0.001, \, 0.01, \, 0.1\right\rbrace$~V/\AA\ and frequencies $\omega /2 \pi \in \, \left\lbrace 2, \, 4, \, 8 \right\rbrace \times 10^{\left\lbrace -1, \, 0, \, 1 \right\rbrace}$~GHz. While the largest field considered is beyond what could be sustained by real water, it remains interesting to include it in the discussion of the various regimes that may arise upon increasing the external driving. For the results presented in Fig.~\ref{fig:SigmaEandOmega}, the magnitudes and frequencies are $E_0 = 2^{p} \, 10^{-2}$~V/\AA\ with $p \in \left\lbrace -3,\,-2, \,..., 9, \, 10 \right\rbrace$ and $\omega / 2 \pi \in  \, \left\lbrace 1, \, 2, \, 4 \right\rbrace \times 10^{\left\lbrace -1, \, 0, \, 1, \, 2\right\rbrace}$~GHz. In such nonequilibrium simulations, we monitor the nonequilibrium current $J_\mathrm{el}(t)$ in the stationary regime, which is reached after a transient regime that depends on $E_0$ and $\omega$ but is shorter than 10 periods $T=2\pi/\omega$ of the field in all cases. We therefore discard the first 10 periods of the signal and the results are averaged over the $T_\mathrm{tot}\omega/2\pi-10$ remaining periods (90 periods for the smallest frequency).

%%%%%%%%%%%%%%%%%%%%%%%%%%%%%%%%%%%%%%%%%%%%%%%%%%%%%%%%%%%%%%%%%%%%%%%%%%%%%%%%%%%%
\section{Results}
\label{sec:Results}

We first discuss generic features of the field- and frequency-dependence of the response to an applied electric field for a confined electrolyte, for ideal particles confined between repulsive walls, with a fixed distance $L$ between the latter in Section~\ref{sec:Results:EandOmega}. We then consider the effects of confinement and adsorption on the walls in Sections~\ref{sec:Results:Confinement} and ~\ref{sec:Results:Adsorption}, respectively. Finally, we discuss the effects of ion-ion interactions in Section~\ref{sec:Results:Interactions}.

%%%%%%%%%%%%%%%%%%%%%%%%%%%%%%%%%%%%%%%%%%
\subsection{Ideal particles: field- and frequency-dependent response}
\label{sec:Results:EandOmega}

\begin{figure}[ht!]
    \begin{center}
    \includegraphics[width=0.45\textwidth]{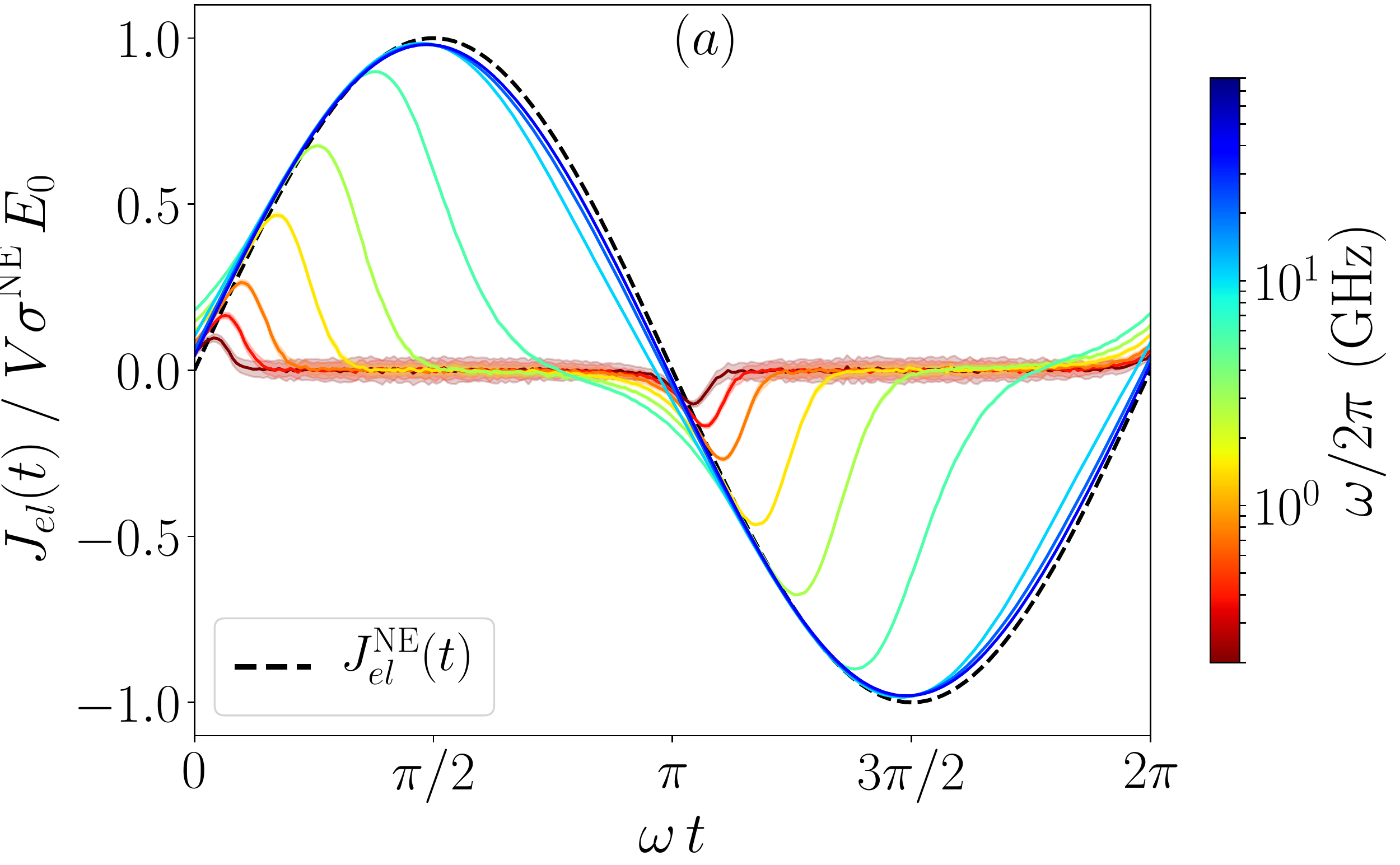} \\
    \vspace{0.2cm}
    \includegraphics[width=0.45\textwidth]{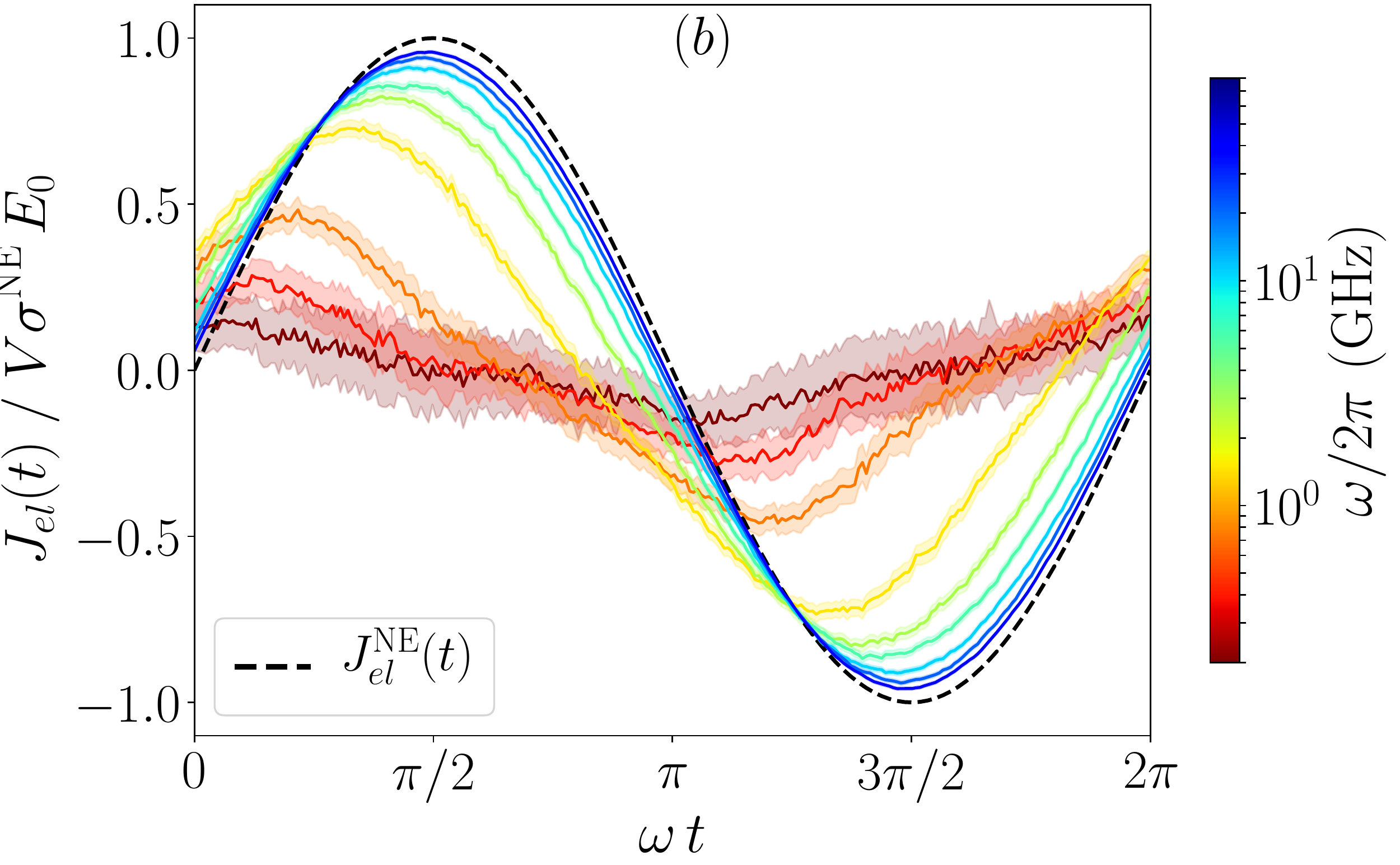} \\
    \vspace{0.2cm}
    \includegraphics[width=0.45\textwidth]{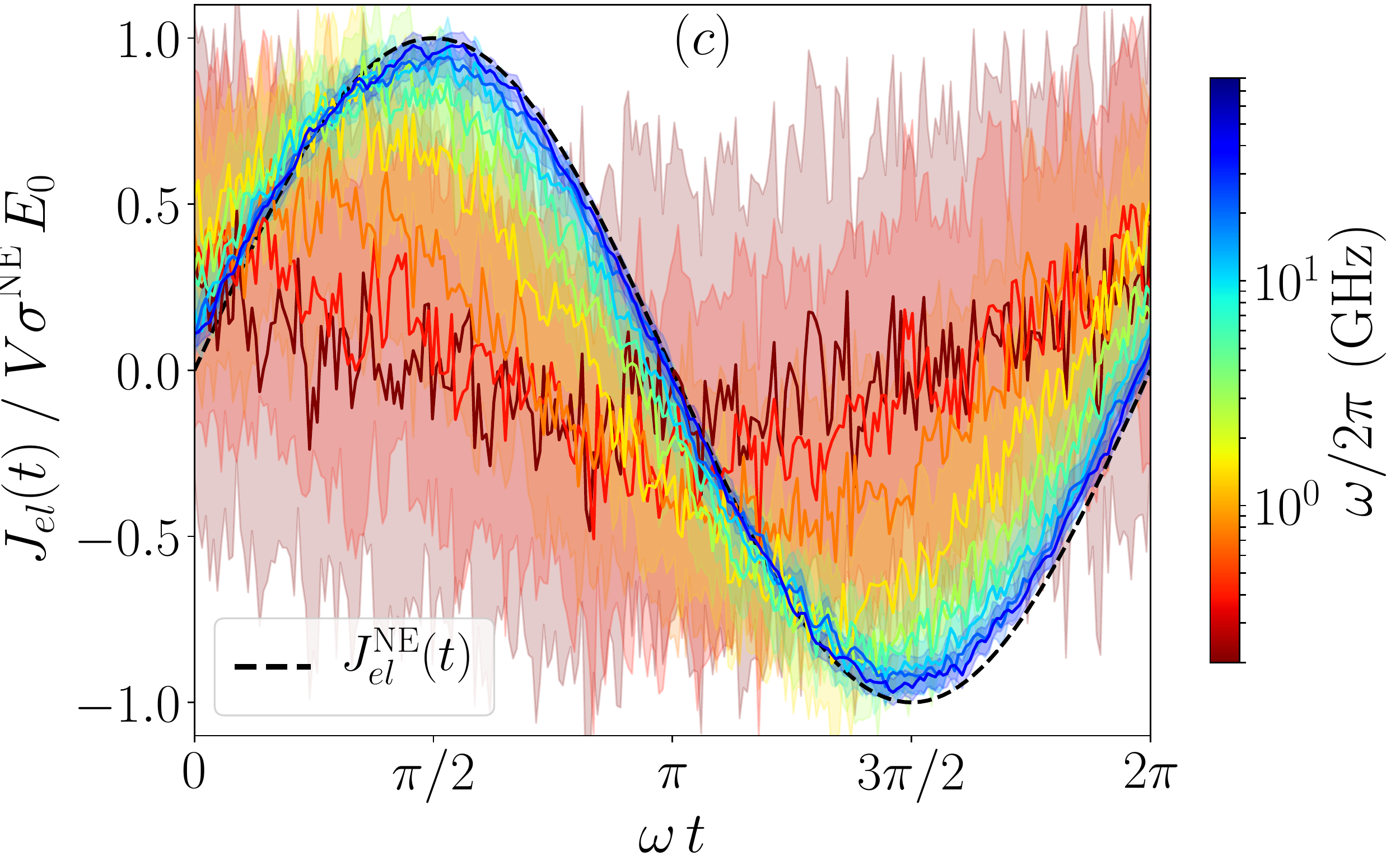}
    \end{center}
    \caption{
    Electric current in the stationary regime of nonequilibrium simulations, for an ideal electrolyte confined between repulsive walls separated by a distance $L=20$~\AA\ (system~I in Table~\ref{tab:systems}), and various magnitudes of the applied field ($E_0 = 10^{-1}$, $10^{-2}$ and $10^{-3}$~V/\AA\ in panels a, b and c, respectively) and frequencies $\omega$ (see colorbars). Results are shown for a single period as a function of $\omega t$, normalized by the maximal current for ideal particles in the absence of confinement by the walls (bulk case), $V\sigma^{\mathrm{NE}}E_0$. Results for the time-dependent current in that case, $J_\mathrm{el}^\mathrm{NE}(t)= V\sigma^{\mathrm{NE}}E(t)$, are also shown as dashed lines. The shaded areas indicate 95\% confidence intervals.
    }
    \label{fig:NEQ_currents}
\end{figure}

We first consider the simplest case of ideal particles confined between repulsive walls (system~I in Table~\ref{tab:systems}) for a fixed distance $L=20$~\AA. Fig.~\ref{fig:NEQ_currents} shows the electric current $J_\mathrm{el}(t)$ defined in Eq.~\eqref{eq:JelBD}, in the stationary regime of nonequilibrium BD simulations with various magnitudes $E_0$ and frequencies $\omega$ of the applied electric field (see Eq.~\eqref{eq:Eoft}). Results are shown for 3 values of $E_0$ as a function of $\omega t$ to compare the currents over a single period, and normalized by the maximal current for ideal particles in the absence of confinement by the walls (bulk case), $V\sigma^{\mathrm{NE}}E_0$. The figure also shows (dashed lines) the time-dependent current in that case, $J_\mathrm{el}^\mathrm{NE}(t)= V \sigma^{\mathrm{NE}}E(t)$.

Let us first consider panel~\ref{fig:NEQ_currents}a, for the largest considered field ($E_0 = 10^{-1}$~V/\AA). For high frequencies, the current almost exactly follows the result corresponding to unconfined electrolytes. For low frequencies, the current first follows the same trend, but then reaches an extremum and decays to zero within each half-period. Upon decreasing the frequency, the maximum current $J_{\max}$ decreases and this maximum is reached after a characteristic time $t_{\max}$ occuring earlier within the period (see Appendix~\ref{sec:Appendix:NL} for a more detailed discussion). As the magnitude of the field decreases (panels~\ref{fig:NEQ_currents}b and~\ref{fig:NEQ_currents}c for $E_0 = 10^{-2}$ and $10^{-3}$~V/\AA, respectively), for a given frequency, the deviations from the ideal response $J_\mathrm{el}^\mathrm{NE}(t)= V\sigma^{\mathrm{NE}}E(t)$ are less pronounced, and the uncertainty on the stationary current increases (for the fixed trajectory length considered here). The latter observation indicates that it is difficult to obtain the frequency-dependent conductivity of the system by considering numerically the limit $E_0\to0$ in Eq.~\eqref{eq:LinearResponse}.

From the definition Eq.~\eqref{eq:JelBD} of $J_\mathrm{el}(t)$, it immediately follows that the deviations from the ideal current are due to the forces experienced by the particles other than the effect of the applied electric field, which in the present case are limited to the short-range repulsive forces exerted by the confining walls. One should then consider two time scales, corresponding to the diffusion and migration of the particles over the relevant characteristic distance between the walls ($\tilde{L}$, discussed below), namely
\begin{equation}
    \tau_{\rm diff} = \frac{\tilde{L}^{2}}{\pi^2 D}
    \label{eq:tauDiff}
\end{equation}
and
\begin{equation}
    \tau_{E} = \frac{\tilde{L}}{\beta e D E_0}
    \, 
    \label{eq:tauE}
\end{equation}
where for simplicity we consider the same diffusion coefficient $D$ for cations and anions, with charges $q_i=\pm e$.

\begin{figure}[ht!]
    \begin{center}
        \includegraphics[width=0.45\textwidth]{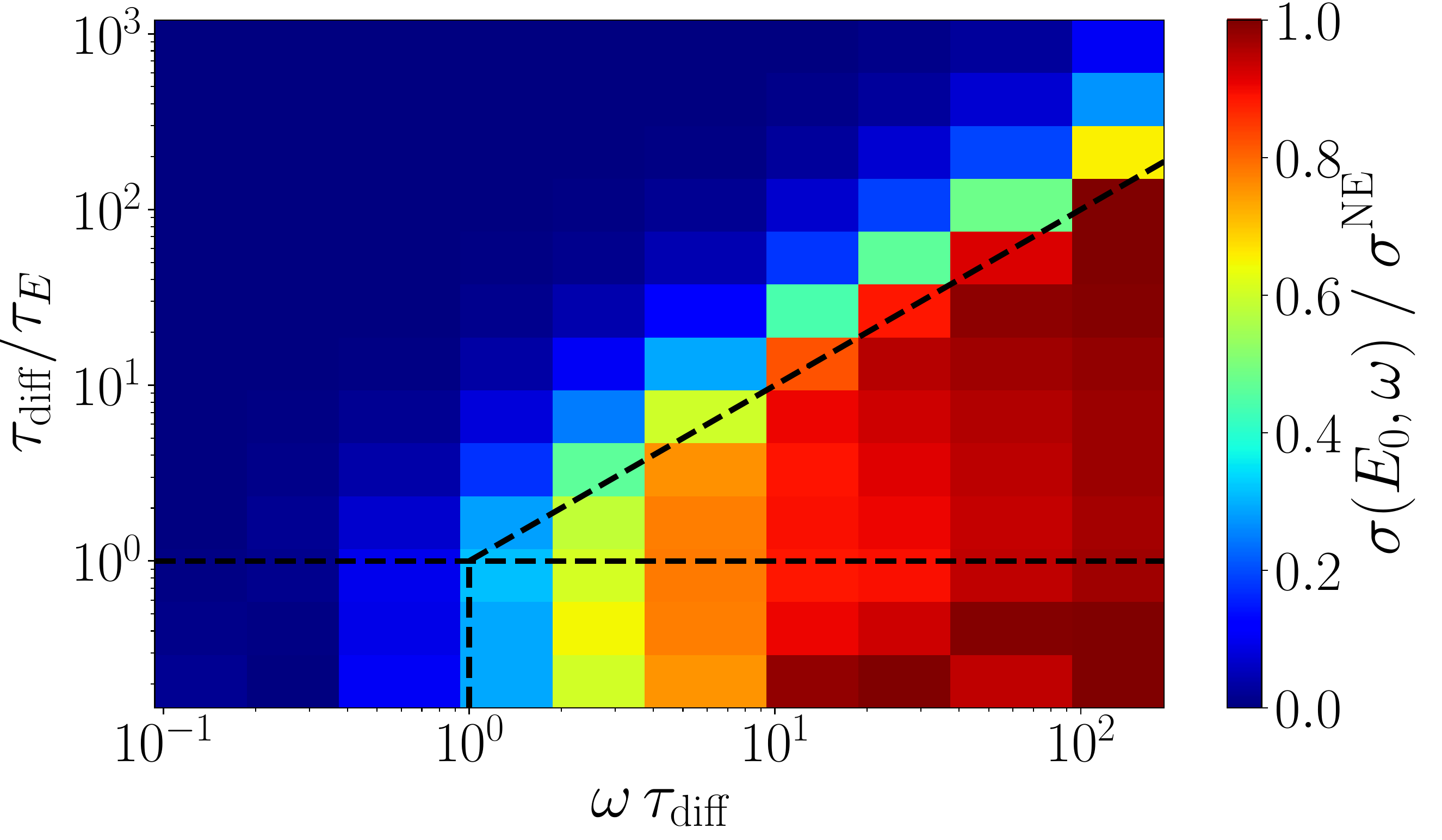}
    \end{center}
    \caption{Field- and frequency-dependent conductivity $\sigma(E_0,\omega)$ (see Eq.~\eqref{eq:SigmaEandOmega}) for ideal particles confined between repulsive walls separated by a distance $L=20$~\AA. The results are shown normalized by the ideal conductivity $\sigma^{\mathrm{NE}}$ as a function of the frequency $\omega$, scaled by the diffusion time, and of the ratio $\tau_{\rm diff}/\tau_{E}$ between the diffusion and migration times (see Eqs.~\eqref{eq:tauDiff} and~\eqref{eq:tauE}).
    The vertical ($\omega\tau_{\rm diff}=1$) and diagonal ($\omega\tau_{E}=1$) dashed lines show the transition between the confined and bulk-like regimes (with vanishing and bulk conductivity, respectively), while the horizontal line ($\tau_{\rm diff}/\tau_{E}=1$) marks the transition between the diffusion- and migration-dominated regimes.
    }
    \label{fig:SigmaEandOmega}
\end{figure}

A systematic investigation of the deviations from the ideal response is shown in the frequency domain in Fig.~\ref{fig:SigmaEandOmega}, which reports the field- and frequency-dependent conductivity $\sigma(E_0,\omega)$ (see Eq.~\eqref{eq:SigmaEandOmega}) for ideal particles confined between repulsive walls separated by a distance $L=20$~\AA. The ideal behavior, corresponding to $\sigma(E_0,\omega)\approx \sigma^{\mathrm{NE}}$, is always recovered at high frequency. In contrast, at low frequency the effect of the confining walls manifests itself as $\sigma(E_0,\omega)\approx 0$. The transition between the two regimes depends on the magnitude of the applied field. For sufficiently large fields, the migration of the ions over the pore width occurs fast with respect to their diffusion ($\tau_E \ll\tau_{\rm diff}$) and deviations of the ideal response are observed when a significant fraction of the particles reaches the walls within the period, \textit{i.e} $\omega \tau_E < 1$. In the opposite limit where the motion of ions is dominated by diffusion, the transition occurs when the latter process is sufficiently fast to homogenize the system within a period of the field, \textit{i.e} $\omega \tau_{\rm diff} < 1$. Further discussion of the nonlinear response in the high-field regime is provided in Appendix~\ref{sec:Appendix:NL}, which confirms that the main origin of the nonlinearity is the fact that the ions reach the walls within the (half-)period of the applied field.

\begin{figure}[ht!]
    \begin{center}
        \includegraphics[width = 0.45\textwidth]{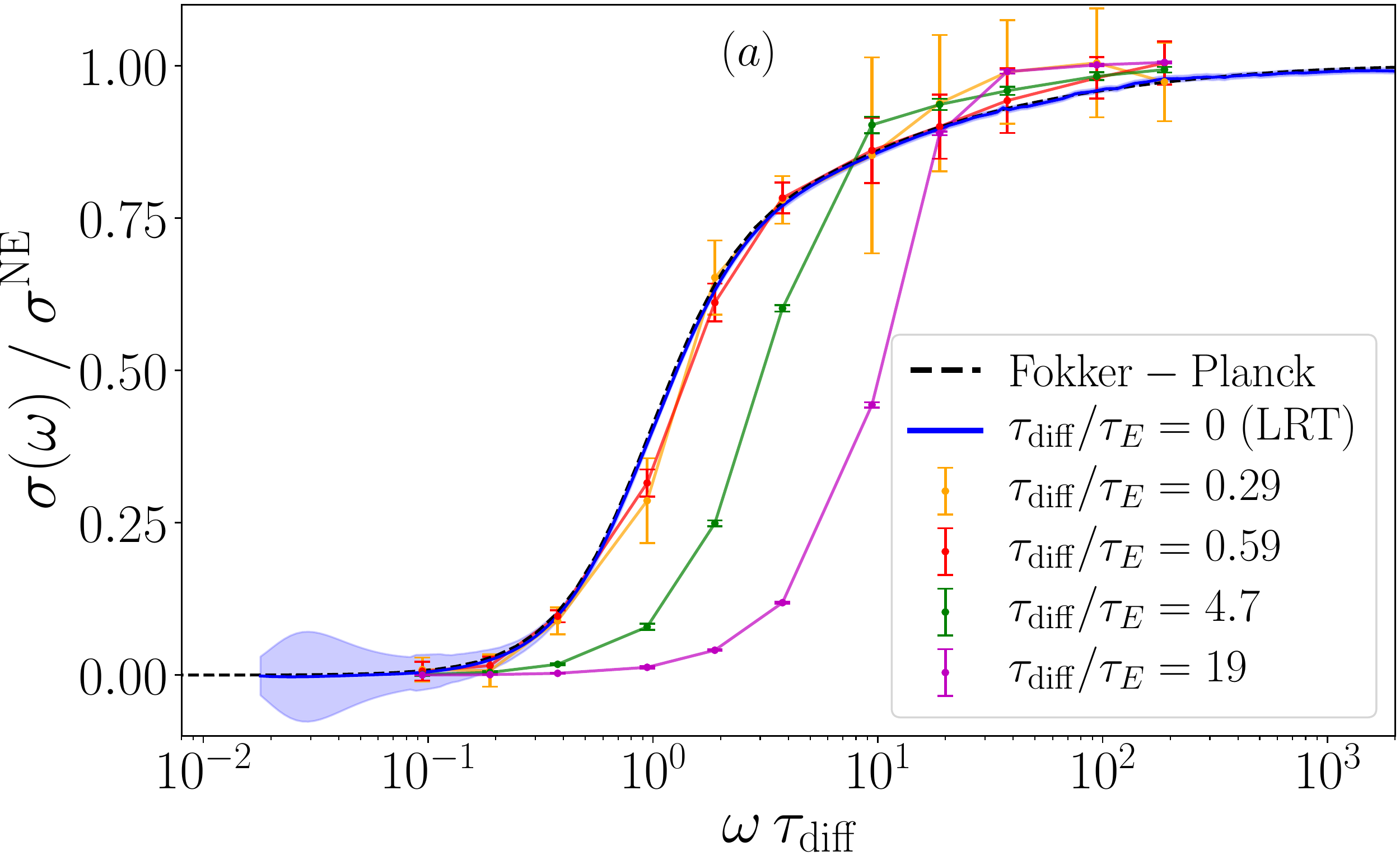}
        \includegraphics[width = 0.45\textwidth]{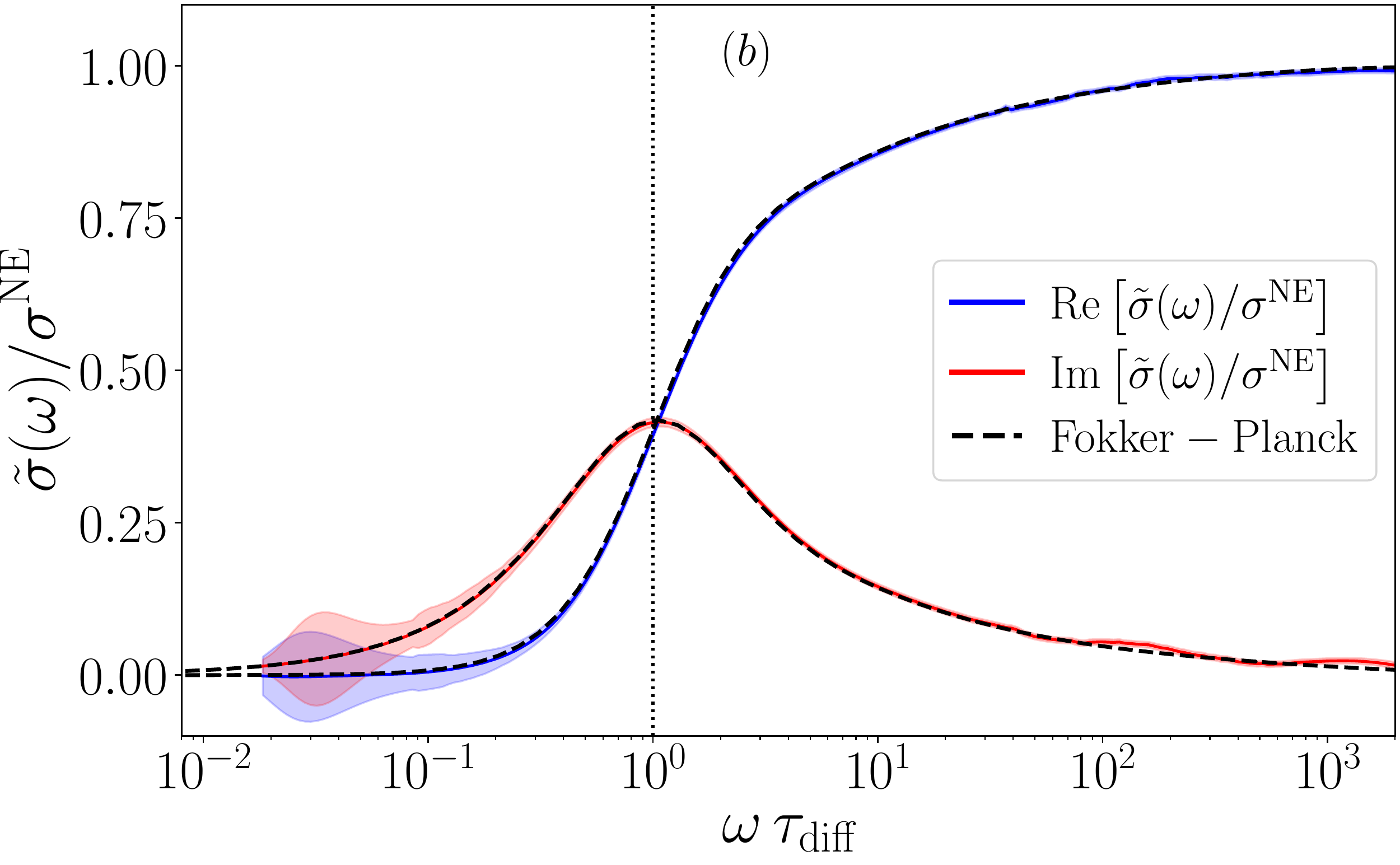}
    \end{center}
    \caption{
    Frequency-dependent conductivity $\sigma(\omega)$ (see Eq.~\eqref{eq:LinearResponse}) for ideal particles confined between repulsive walls separated by a distance $L=20$~\AA. The results are shown normalized by the ideal conductivity $\sigma^{\mathrm{NE}}$ as a function of the frequency $\omega$, scaled by the diffusion time $\tau_{\rm diff}$ (see Eq.~\eqref{eq:tauDiff}). (a) Results from equilibrium Brownian dynamics simulations (blue line) using Eq.~\eqref{eq:GK_conductivity_BD} within linear response theory (LRT) are compared to $\sigma(E_0,\omega)$ from nonequilibrium simulations with several fields (corresponding to various values of $\tau_{\rm diff}/\tau_{E}$, indicated by the color) and selected frequencies (symbols). The shaded area and the error bars indicate 95\% confidence intervals. Results obtained by solving the Fokker--Planck equation in the absence of an applied field (see Appendix~\ref{sec:Appendix:FP}), are also shown (dashed lines).
    (b) Real (blue) and imaginary (red) parts of the frequency-dependent conductivity from equilibrium Brownian dynamics simulations (solid lines) and from the Fokker--Planck equation (dashed lines). The vertical dotted line indicates $\omega\tau_{\rm diff}=1$.
    }
    \label{fig:neq_vs_eq}
\end{figure}

We now investigate in more detail the linear response obtained in the limit of vanishing fields (see Eq.~\eqref{eq:LinearResponse}), \textit{i.e} $\tau_{\rm diff} / \tau_E \to 0$, and characterized by the (field-independent and) frequency-dependent conductivity $\sigma(\omega)$. Fig.~\ref{fig:neq_vs_eq}a compares the results obtained over the whole frequency spectrum from equilibrium BD simulations, using Eq.~\eqref{eq:GK_conductivity_BD}, for ideal particles confined between repulsive walls separated by a distance $L=20$~\AA, to $\sigma(E_0,\omega)$ from nonequilibrium simulations with several fields, corresponding to various values of $\tau_{\rm diff}/\tau_{E}$, and selected frequencies, as well as to the results obtained by solving the Fokker-Planck equation in the absence of an applied field (see Appendix~\ref{sec:Appendix:FP}). 

The predictions from equilibrium simulations using linear response theory (blue line) are in excellent agreement with the solution of the FP equation in the absence of applied field (dashed line) over the whole frequency range. This validates the use of the Green--Kubo expression for BD simulations, Eq.~\eqref{eq:GK_conductivity_BD}. The uncertainty obtained with this method grows at lower frequencies, due to the finite length of the trajectories. Both results are also consistent with the results from nonequilibrium simulations with small applied fields, as expected. However, the latter have to be determined separately for each frequency, whereas the equilibrium routes provides $\sigma(\omega)$ for all frequencies from a single simulation. In addition, the uncertainty on the results grows as the magnitude of the field decreases (as already observed for the stationary current in Fig.~\ref{fig:NEQ_currents}). Finally, we note that the nonequilibrium route to $\sigma(\omega)$ requires simulations with several fields to ensure that one is in the linear response regime. It allows considering the response of the system beyond this regime, as illustrated with the results for the larger fields. The latter show that the transition between the confined regime at low frequency and the ideal response at high frequency is shifted towards higher frequencies upon increasing the magnitude of the field, as already discussed in Fig.~\ref{fig:SigmaEandOmega}.

Fig.~\ref{fig:neq_vs_eq}b then shows both the real (as in panel Fig.~\ref{fig:neq_vs_eq}a) and imaginary parts of the frequency-dependent conductivity. The results from equilibrium Brownian dynamics simulations are in good agreement with the solution of the Fokker--Planck equation also for the imaginary part. The latter displays a maximum for $\omega\tau_{\rm diff}=1$, pointing to the role of this diffusive time scale, which will be further discussed in Section~\ref{sec:Results:Confinement}. As mentioned when introducing the complex conductivity (see Eq.~\ref{eq:LinearResponse}), all the relevant information is encoded in either the real or imaginary part, since they are related by the Kramers-Kronig relations. However the imaginary part provides a more convenient way to define the characteristic crossover frequency as:
\begin{equation}
  \omega^* = \argmax_{\omega}\;  \operatorname{Im}\left[ \tilde{\sigma}(\omega) \right]
  \; .
  \label{eq:DefOmegaStar}
\end{equation}
In the case shown in Fig.~\ref{fig:neq_vs_eq}b, the real and imaginary parts are approximately equal to $\approx 0.4\sigma_{\rm NE}$ for this frequency.
As a final remark, we note that the plateau at high-frequency of the real part of the conductivity, $\lim_{\omega\to\infty} \sigma(\omega)=\sigma^{\mathrm{NE}}$, is a consequence of the Brownian description Eq.~\ref{eq:BD}, which neglects the relaxation of velocities at short times, as discussed in Section~\ref{sec:BD:Model}. With a more realistic description of these timescales, using underdamped Langevin dynamics or molecular dynamics, one would recover the expected decay to zero of the conductivity in the limit $\omega\to\infty$ where ions cannot follow the applied field.

%%%%%%%%%%%%%%%%%%%%%%%%%%%%%%%%%%%%%%%%%%
\subsection{Ideal particles: effect of confinement}
\label{sec:Results:Confinement}

\begin{figure}[ht!]
     \centering
     \includegraphics[width=0.45\textwidth]{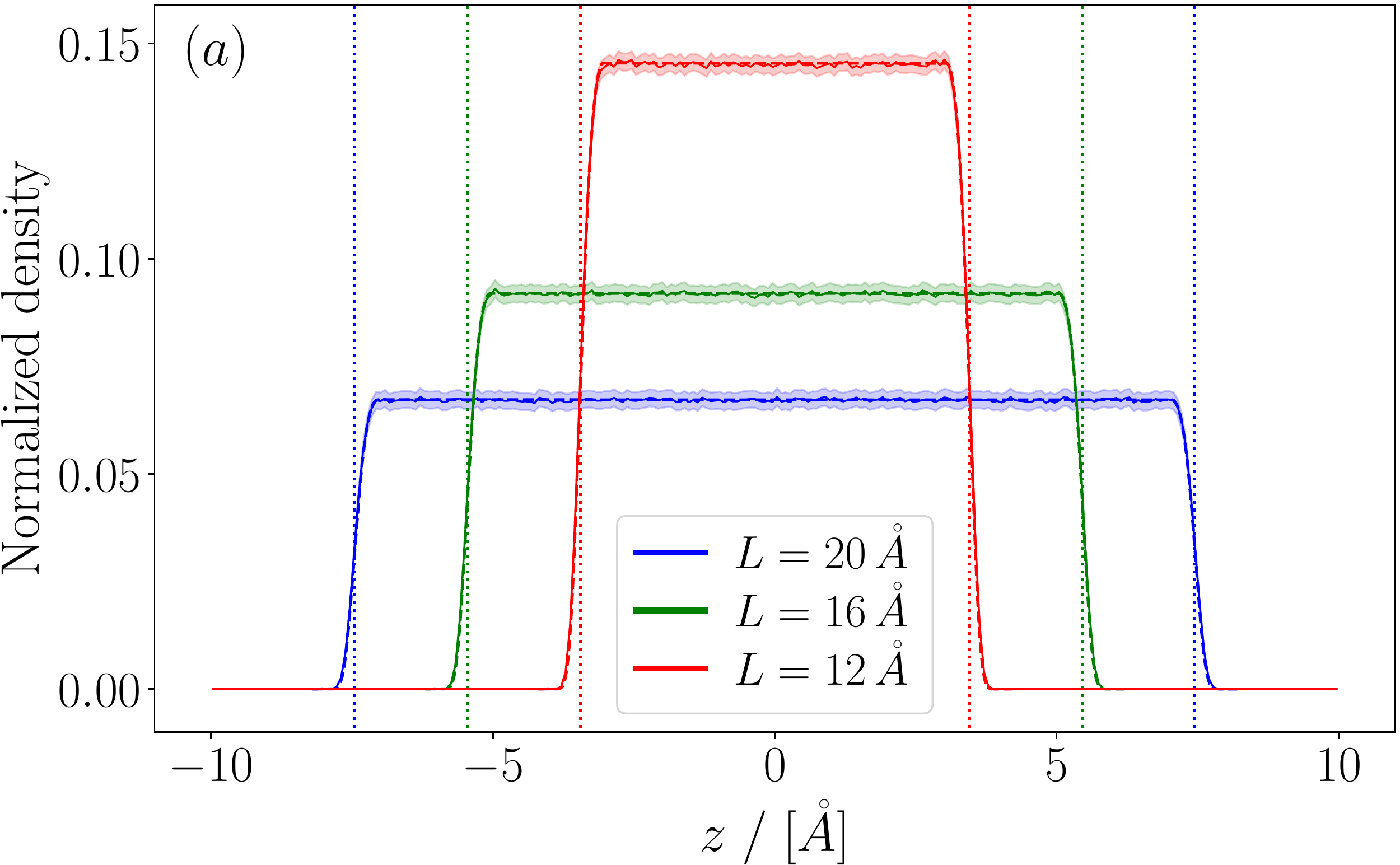} \\
     \includegraphics[width=0.45\textwidth]{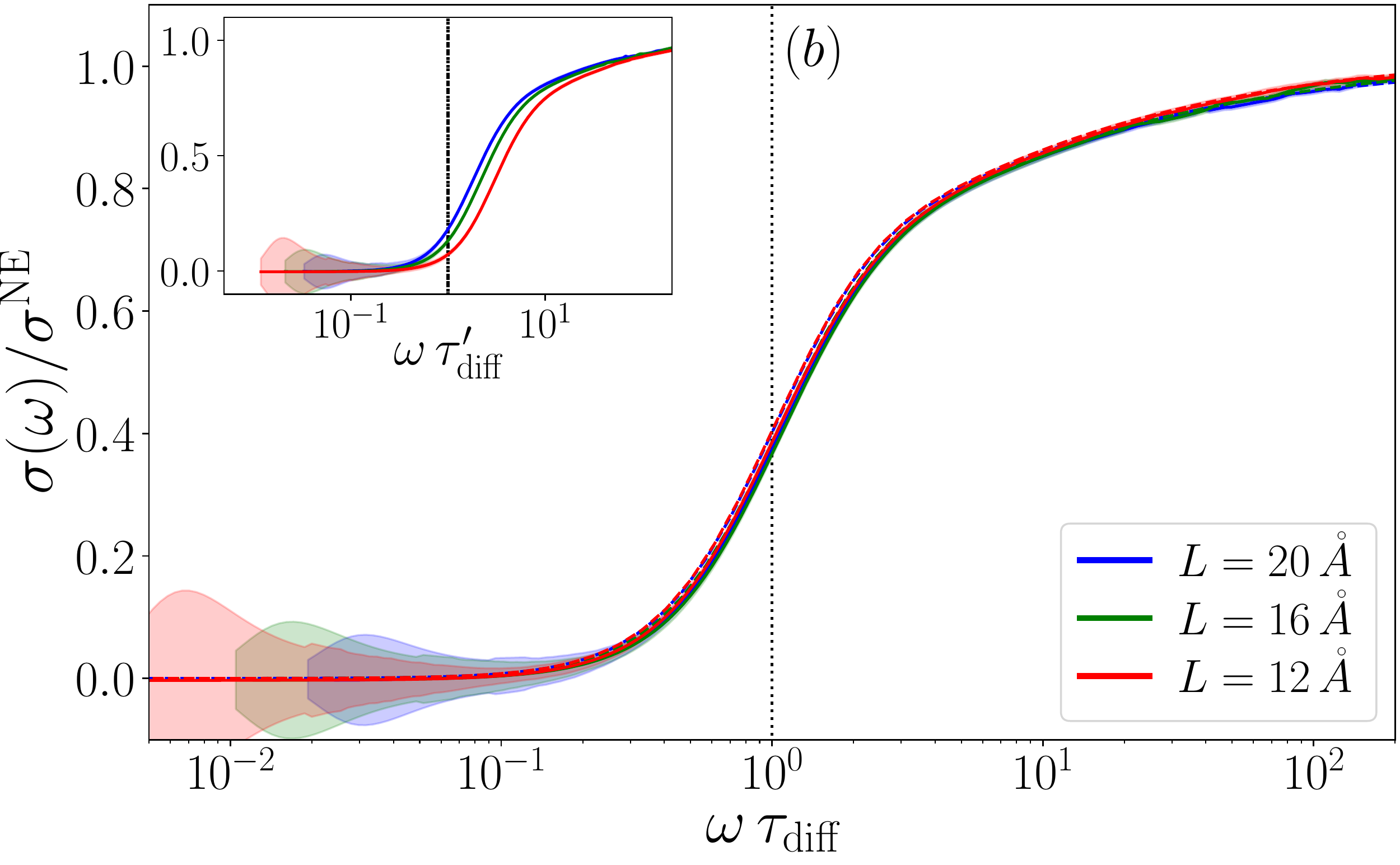} \\
    \caption{
            (a) Normalized equilibrium density profiles for 3 distances $L$ between the repulsive walls. The simulation results (solid lines, with shaded areas indicating 95\% confidence intervals, follow the Boltzmann distribution expected for ideal particles (dashed lines). The vertical dotted lines indicate the positions of the Gibbs Dividing Surfaces (see Eq.~\eqref{eq:GDS}); the distance between the latter defines the characteristic length $\tilde{L}$.
            (b) Frequency-dependent conductivity $\sigma(\omega)$ for the 3 distances $L$. The results are shown normalized by the ideal conductivity $\sigma^{\mathrm{NE}}$ as a function of the frequency $\omega$, scaled by the diffusion time $\tau_{\rm diff}$ defined from $\tilde{L}$ in Eq.~\eqref{eq:tauDiff}. The inset shows the same quantity when defining the diffusion time $\tau_{\rm diff}'$ using $L$ instead of $\tilde{L}$.
            }
    \label{fig:sigmaL}
\end{figure}

All the results presented so far correspond to a fixed distance $L=20$~\AA\ between the repulsive walls. We now consider the effect of this distance on the frequency-dependent conductivity and discuss the characteristic length $\tilde{L}$ introduced to define the diffusion and migration time scales in Eqs.~\eqref{eq:tauDiff} and~\eqref{eq:tauE}. Fig.~\ref{fig:sigmaL}a shows the normalized equilibrium density profiles for 3 distances $L$ between the repulsive walls. As expected for non-interacting particles, they follow the Boltzmann distribution $\rho(z)\propto \mathrm{e}^{-\beta U(z)}$. In the present case of short-range particle-wall repulsion, this leads to a flat profile except near the wall, where the density decays to zero over a typical distance $\sim\sigma_\mathrm{w}$ (see Eqs.~\eqref{eq:Walls:Vads} and~\eqref{eq:Walls:Vrep}). For such smooth density profiles, an equivalent sharp interface between a homogeneous region with bulk density $\rho_\mathrm{b}$ and an empty region inside the wall can be defined as the Gibbs Dividing Surface (GDS), at the position $z_\mathrm{GDS}^-$ such that (for the left interface in the region $z<0$, and a similar definition of $z_\mathrm{GDS}^+$ for the other interface):
\begin{equation}
    \int_{-\infty}^{z_\mathrm{GDS}^-} \rho(z) \, {\rm d}z = \int_{z_\mathrm{GDS}^-}^0 \left[ \rho_\mathrm{b} - \rho(z) \right] \, {\rm d}z
    \, .
    \label{eq:GDS}
\end{equation}
These positions are indicated as vertical dotted lines in Fig.~\ref{fig:sigmaL}a. The width of the region occupied by the fluid, which defines the characteristic confining length is then the distance between these two interfaces, \textit{i.e.}
\begin{equation}
    \tilde{L} = z_\mathrm{GDS}^+ - z_\mathrm{GDS}^-
    \, .
    \label{eq:tildeL}
\end{equation}
In practice, we find numerically that for the considered repulsive wall this is well approximated by $\tilde{L} \approx L- 1.7\sigma_\mathrm{w}$. In order to illustrate the relevance of Eq.~\eqref{eq:tildeL} to define the characteristic times in Eqs.~\eqref{eq:tauDiff} and~\eqref{eq:tauE}, we show in panel~\ref{fig:sigmaL}b the frequency-dependent conductivity for 3 distances between walls, as a function of the frequency scaled by $\tau_{\rm diff}$. This scaling perfectly captures the effect of confinement, while it is not the case if $L$ is used instead of $\tilde{L}$ to define the diffusion time, as illustrated in the inset.

%%%%%%%%%%%%%%%%%%%%%%%%%%%%%%%%%%%%%%%%%%
\subsection{Ideal particles: effect of adsorption}
\label{sec:Results:Adsorption}

\begin{figure}[ht!]
    \begin{center}
    \includegraphics[width=0.44\textwidth]{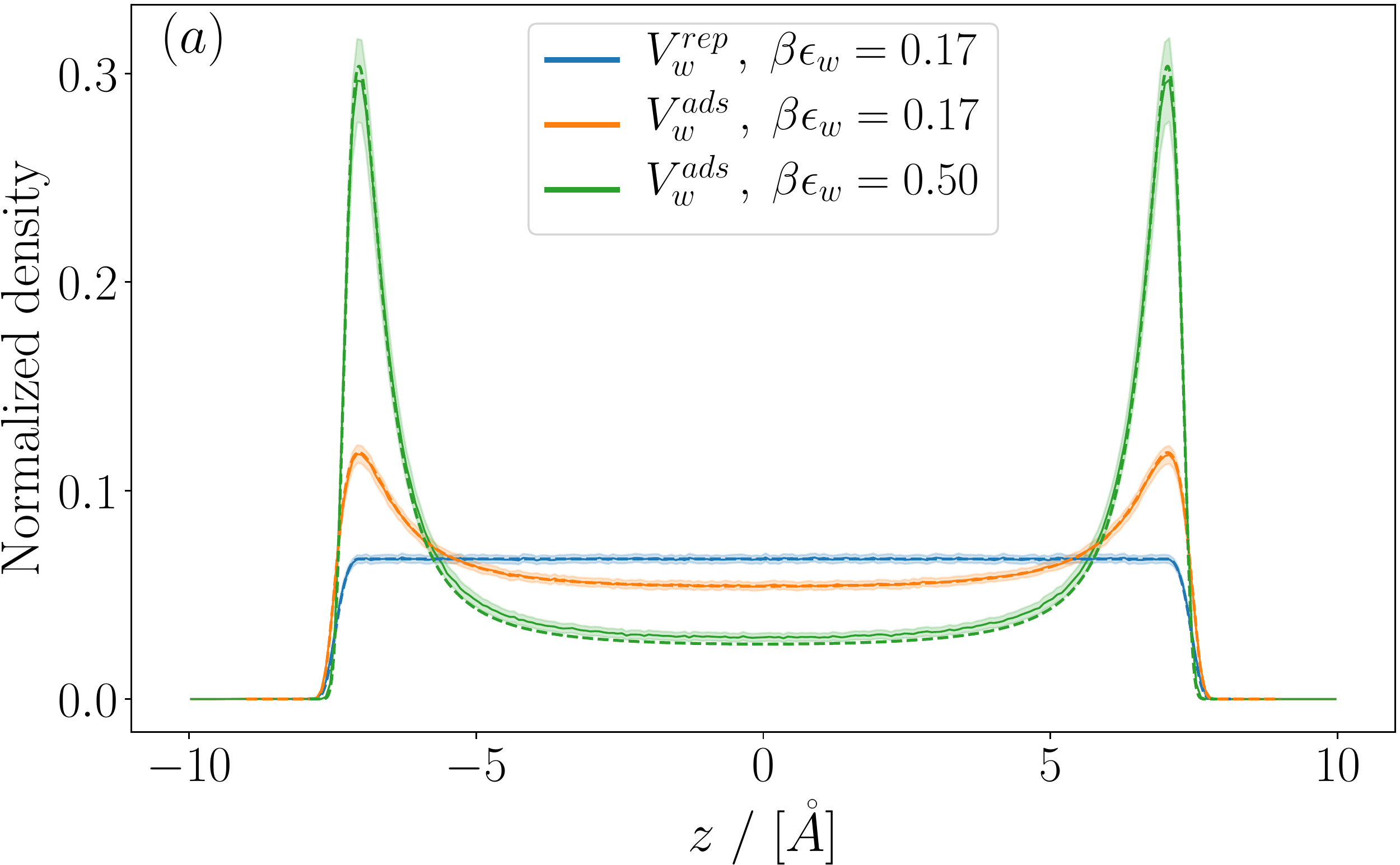} 
    \\ \vspace{0.1cm}
    \includegraphics[width=0.45\textwidth]{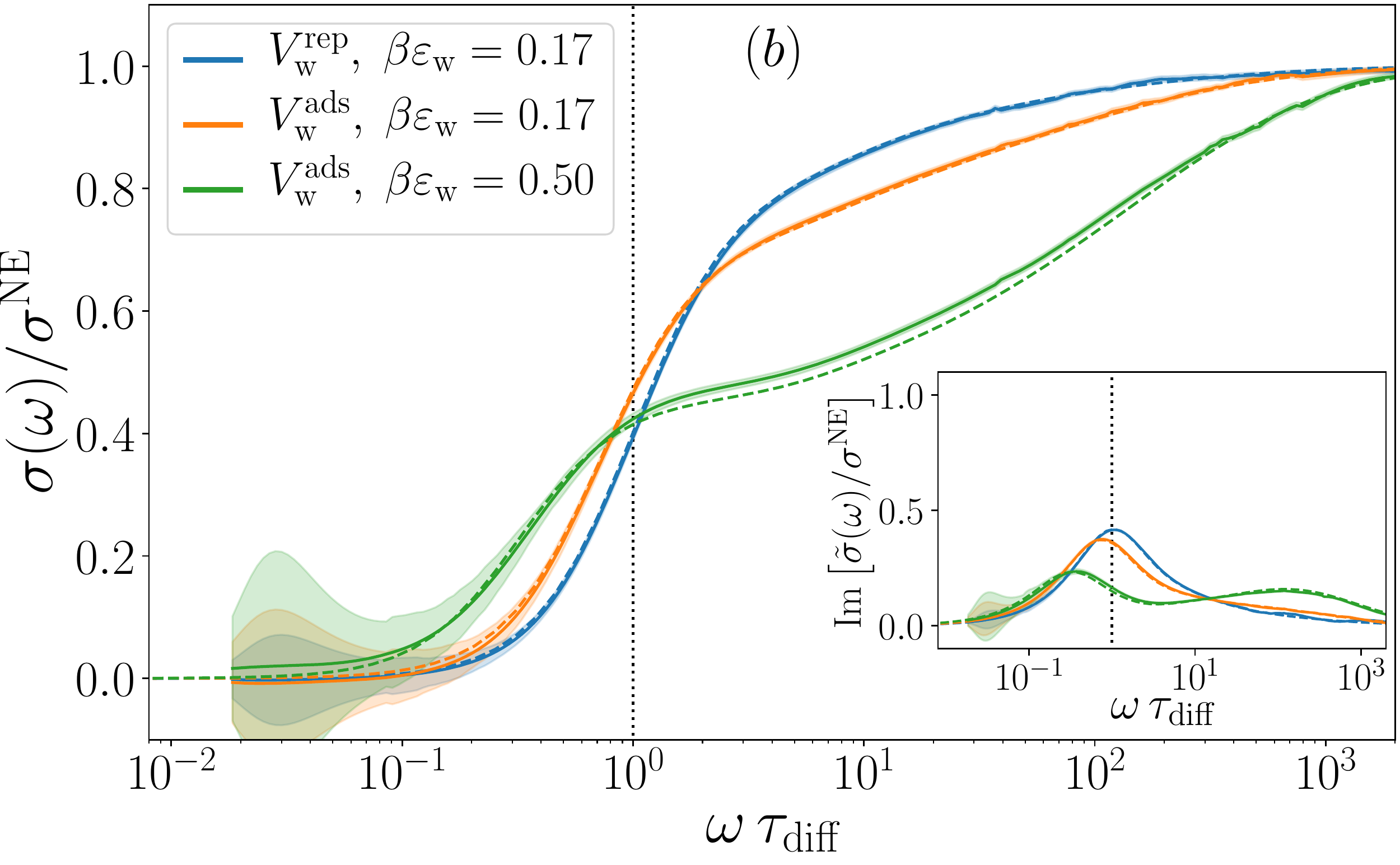}
    \end{center}
    \caption{
    (a) Normalized equilibrium density profiles for a fixed distance $L=20$~\AA\ between the walls and three particle-wall interactions (see Eqs.~\eqref{eq:Walls:Vads} and~\eqref{eq:Walls:Vrep}): repulsive with $\beta \epsilon_\mathrm{w} = 0.17$, attractive with $\beta \epsilon_\mathrm{w} = 0.17$ and attractive with $\beta \epsilon_\mathrm{w} = 0.50$. The simulation results (solid lines, with shaded areas indicating 95\% confidence intervals), follow the Boltzmann distribution expected for ideal particles (dashed lines). 
    (b) Frequency-dependent conductivity $\sigma(\omega)$ for the same systems, normalized by the ideal conductivity $\sigma^{\mathrm{NE}}$, as a function of the frequency $\omega$, scaled by the diffusion time $\tau_{\rm diff}$ defined (see Eq.~\eqref{eq:tauDiff}) with $\tilde{L}$ in the absence of adsorption.
    The simulation results (solid lines), are in good agreement with the numerical solution of the Fokker--Planck equation for ideal particles (dashed lines). 
    The inset in panel (b) shows the imaginary part of the frequency-dependent conductivity.
    }
    \label{fig:adsorption}
\end{figure}

The effect of adsorption on the frequency-dependent conductivity is explored by considering the 3 particle-wall interaction potentials illustrated in Fig.~\ref{fig:potentials}, for a fixed distance between the walls $L=20$~\AA: repulsive with $\beta \epsilon_\mathrm{w} = 0.17$, attractive with $\beta \epsilon_\mathrm{w} = 0.17$ and attractive with $\beta \epsilon_\mathrm{w} = 0.50$ (see Eqs.~\eqref{eq:Walls:Vads} and~\eqref{eq:Walls:Vrep}). The first case corresponds to the one discussed in sections~\ref{sec:Results:EandOmega} and~\ref{sec:Results:Confinement} and serves as a reference. Fig.~\ref{fig:adsorption}a shows the normalized equilibrium density profiles, which follow the Boltzmann distribution $\rho(z)\propto \mathrm{e}^{-\beta U(z)}$, as expected for ideal particles. Upon introducing short-range attraction to the wall, one observes the desired adsorption of particles with peaks in the density near the surfaces, which grow with increasing $\beta \epsilon_\mathrm{w}$.

Fig.~\ref{fig:adsorption}b then reports the corresponding frequency-dependent conductivity. The simulation results (solid lines), are in good agreement with the numerical solution of the Fokker--Planck equation for ideal particles (dashed lines). While adsorption does not change the limits of vanishing conductivity for $\omega\to0$ (due to confinement by the walls) and of ideal conductivity for $\omega\to\infty$ (when the driving oscillates too fast to feel the presence of the walls), it does change the cross-over between these limits. On the one hand, the frequency at which $\sigma(\omega)$ starts increasing is shifted towards lower values of~$\omega$. On the other hand, the regime of ideal conductivity is reached for higher frequencies. These observations, which are also visible in the imaginary part of the frequency-dependent conductivity (insert of panel Fig.~\ref{fig:adsorption}b) indicate that adsorption results in both slower and faster equilibrium current fluctuations than with repulsive walls. Without entering into a systematic analysis of the combined effects of the attraction strength $\beta \epsilon_\mathrm{w}$ and the distance between walls $L$, which is out of the scope of the present work, one can identify the faster process as the diffusion within the potential wells at the surface and the slower one as the diffusion between the two potential wells. These results confirm the possibility to use the frequency-dependent conductivity as a probe of adsorption on surfaces -- and the relevance of BD simulations to analyze it.

%%%%%%%%%%%%%%%%%%%%%%%%%%%%%%%%%%%%%%%%%%
\subsection{Electrolytes: effect of ion-ion interactions}
\label{sec:Results:Interactions}

In the previous sections~\ref{sec:Results:EandOmega} to~\ref{sec:Results:Adsorption}, the particles interacted only with the confining walls and the external fields, but not between themselves. A more realistic model of electrolyte solutions requires taking into account the long range electrostatic interactions between ions, as well as short-range repulsion preventing in particular the collapse of oppositely charged ions. Here, we explore the effect of such ion-ion interactions (see Eq.~\eqref{eq:PairPotential}) on the frequency-dependent conductivity $\sigma(\omega)$, and consider the role of the salt concentration $C_s$ and the effective distance $\tilde{L}$ between the walls. 

In the bulk, the mean-field Debye--H\"uckel theory of electrolytes, valid for sufficiently low concentrations, results in the well-known picture of an ionic cloud around each ion.  The electrostatic potential and ionic concentrations decay as $\mathrm{e}^{-r/\lambda_D}/r$, with $r$ the distance from the ion and the Debye screening length
\begin{equation}
    \lambda_D = \left( 4 \pi l_{\rm B} \sum_{\alpha} c_\alpha Z_\alpha^2 \right)^{-1/2}
    \, ,
    \label{eq:lambdaD}
\end{equation}
where the sum runs over species $\alpha$ with concentrations $c_\alpha$ and with the Bjerrum length $l_{\rm B}=e^2/4\pi \epsilon_0\epsilon_r k_{\rm B}T$. In the present case of a 1:1 electrolyte, this reduces to $\lambda_D = 1/\sqrt{8\pi l_{\rm B} C_s}$. In addition, charge fluctuations are predicted to relax over a typical timescale called the Debye time
\begin{equation}
    \tau_{\rm Debye} = \left( 4 \pi l_{\rm B}  \sum_{\alpha} c_\alpha Z_\alpha^2 D_\alpha \right)^{-1}
    \, ,
    \label{eq:tauD}
\end{equation}
which simplifies to $\tau_{\rm Debye}=\lambda_D^2/D$ when cations and anions have the same diffusion coefficient $D$. In the following, we will compare the numerical results to this simpler estimate using the average diffusion coefficient $D=(D_+ + D_-)/2$, \textit{i.e.} neglect the higher-order effect of the transient internal field induced by different diffusivities and leading to the coupled diffusion of both ions (Nernst--Hartley equation)~\cite{robinson_stokes_electrolyte_solutions_book}. While many features of the ionic dynamics are overall preserved for unequal diffusion coefficients (see \emph{e.g.} Ref.~\citenum{palaia2023charging} for an illustration on the charging dynamics of nanocapacitors), dramatic effects have been reported for bulk electrolytes under large AC fields~\cite{hashemi_oscillating_electric_2018}.

The situation is expected to be more complex under confinement, as discussed in particular by Bazant \textit{et al.} for the charging dynamics of a nanocapacitor~\cite{bazant_diffuse-charge_2004}. In the linear response regime and in the limit of thin electric double layers ($\lambda_D\ll \tilde{L}$), they confirmed the previously reported role~\cite{macdonald_double_1970,kornyshev_electric_1977,kornyshev_conductivity_1981} of $\tau_{RC}=\lambda_D \tilde{L}/2D$, which corresponds to the characteristic time of an $RC$ circuit with capacitance predicted by Debye-H\"uckel theory and resistance corresponding to the Nernst-Einstein conductivity (Eq.~\eqref{eq:NE}). They further provided the leading correction in terms of the ratio between the Debye length and inter-electrode distance ($\tilde{L}$ in the present work, instead of $2L$ in Ref.~\citenum{bazant_diffuse-charge_2004}).

\begin{figure}[ht!]
  \centering
     \includegraphics[width=0.45\textwidth]{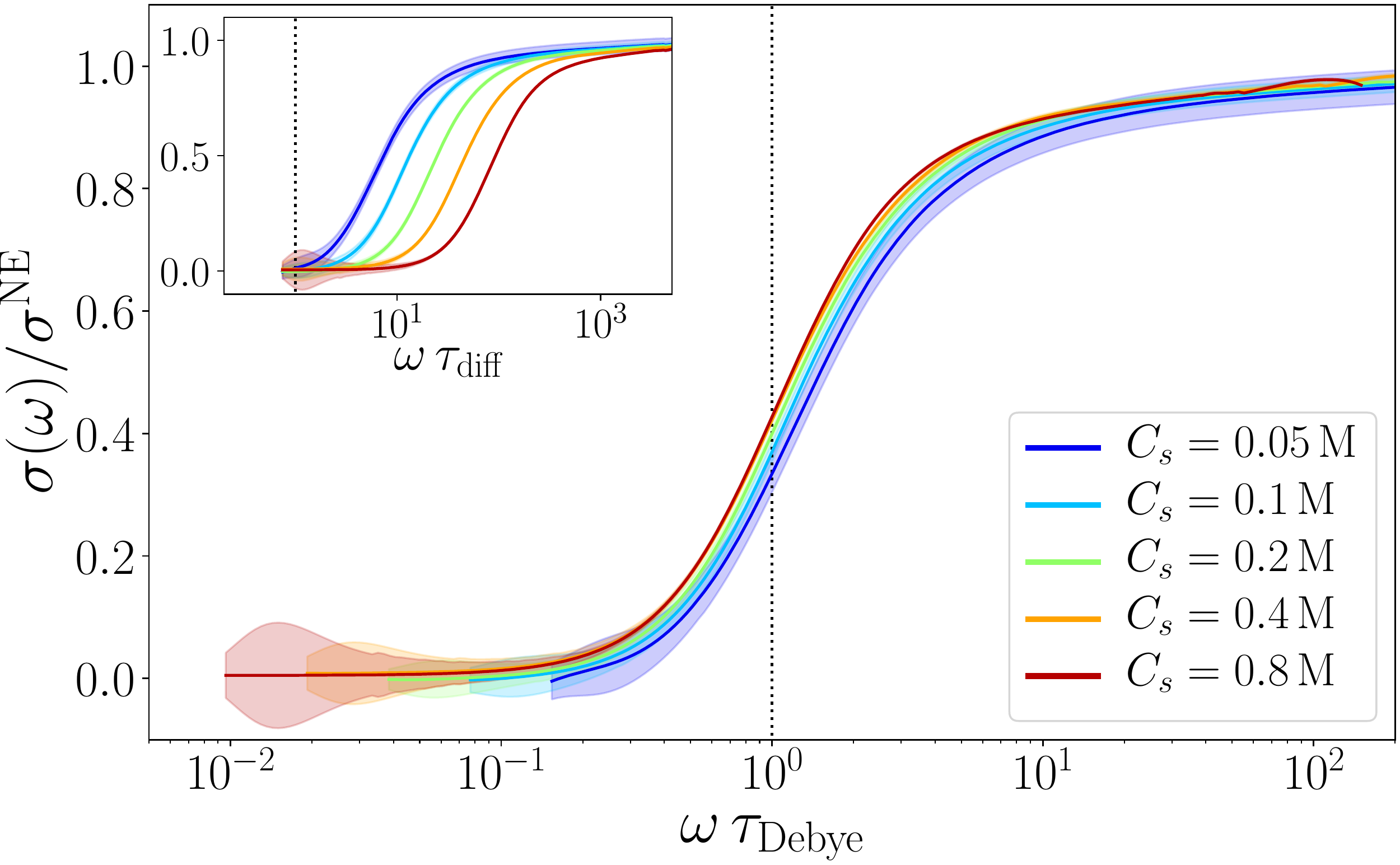} \\
     \caption{
        Frequency-dependent conductivity $\sigma(\omega)$ normalized by the ideal conductivity $\sigma^{\mathrm{NE}}$ for a fixed distance $L=100$~\AA\ between the walls and 5 salt concentrations $C_s$ (see ``Electrolyte I'' in Table~\ref{tab:systems}). These conditions correspond to the limit of thin electric double layers ($\lambda_D \ll \tilde{L}$).
        The frequency is scaled by the Debye time $\tau_{\rm Debye}$ (with the dotted vertical line indicating $\omega\tau_{\rm Debye}=1$). The inset shows the same results as a function of the frequency scaled by the diffusion time $\tau_{\rm diff}$ (with the dotted vertical line indicating $\omega\tau_{\rm diff}=1$).
     }
     \label{fig:sigmaCssmallEDL}
\end{figure}

Fig.~\ref{fig:sigmaCssmallEDL} reports the frequency-dependent conductivity $\sigma(\omega)$ normalized by the ideal conductivity $\sigma^{\mathrm{NE}}$ for a fixed distance $L=100$~\AA\ and 5 salt concentrations $C_s$ (see ``Electrolyte I'' in Table~\ref{tab:systems}), corresponding to the limit of thin electric double layers ($\lambda_D \ll \tilde{L}$). The shape of the curves is similar to the case of ideal particles, with a crossover from a vanishing conductivity regime at low frequency (due to confinement) to an ideal conductivity regime at high frequency. Compared to the previous case of noninteracting particles, the transition occurs for $\omega^*\tau_{\rm Debye}\approx1$ for all concentrations (with the crossover frequency defined by Eq.~\eqref{eq:DefOmegaStar}, even though we do not report the imaginary part in Fig.~\ref{fig:sigmaCssmallEDL}), \textit{i.e.} the relevant time scale is now the Debye relaxation time $\tau_{\rm Debye}$. The latter captures the dependence on the concentration, illustrated in the inset which reports the same results as a function of the frequency scaled by the diffusion time $\tau_{\rm diff}$.

\begin{figure}[ht!]
    \begin{center}
      \includegraphics[width = 0.49\textwidth]{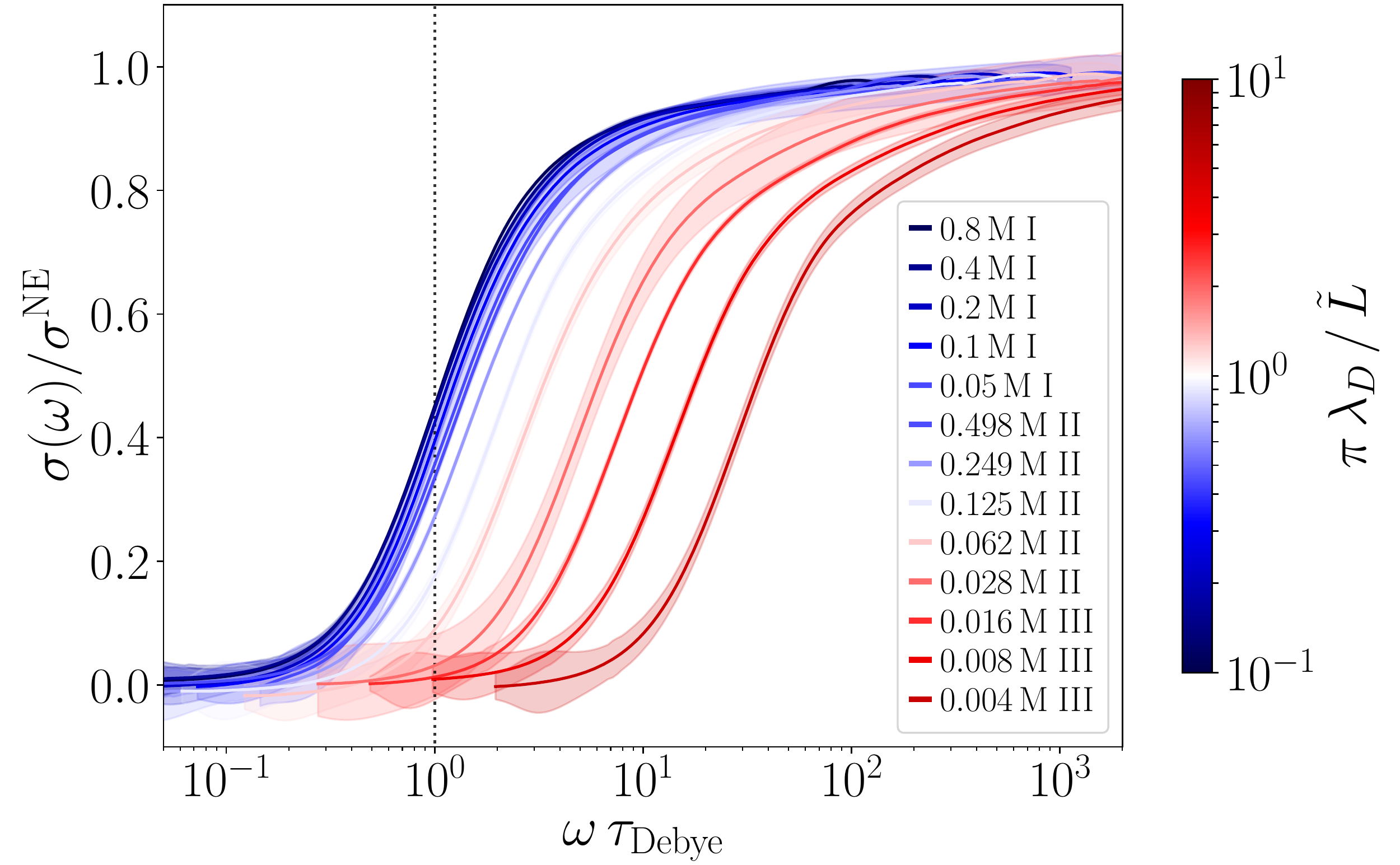}
    \end{center}
    \caption{ 
    Frequency-dependent conductivity $\sigma(\omega)$ normalized by the ideal conductivity $\sigma^{\mathrm{NE}}$ for a wide range of concentrations and several distances between the walls (see Table~\ref{tab:systems} for the definition of the systems labeled I, II and III). The results are shown as a function of the frequency scaled by the Debye time $\tau_{\rm Debye}$ and colored according to the ratio $\pi\lambda_D / \tilde{L}$, with $\lambda_D$ the Debye length. The vertical dotted line indicates $\omega\tau_{\rm Debye}=1$, which corresponds to the crossover between low and high frequency regimes in the limit $\pi\lambda_D / \tilde{L} \ll 1$.
    }
    \label{fig:sigmaAllCsL}
\end{figure}

Fig.~\ref{fig:sigmaAllCsL} summarizes the results for all the considered electrolyte systems (see Table~\ref{tab:systems}) covering a wide range of salt concentrations $C_s$ and several distances $L$ between the walls. The frequency-dependent conductivity is shown as a function of the frequency scaled by the Debye time $\tau_{\rm Debye}$ and colored according to the ratio $\pi\lambda_D / \tilde{L}$, with $\lambda_D$ the Debye length. The results for the systems labeled with I, where $\pi \lambda_D / \tilde{L} \ll 1$, are the same as in Fig.~\ref{fig:sigmaCssmallEDL}, with a transition between the low and high conductivity regimes for $\omega^*\tau_{\rm Debye}\approx 1$. Fig.~\ref{fig:sigmaAllCsL} shows that beyond the limit of thin double layers, the crossover is not simply governed by the Debye relaxation time.

\begin{figure}[ht!]
    \begin{center}
      \includegraphics[width = 0.48\textwidth]{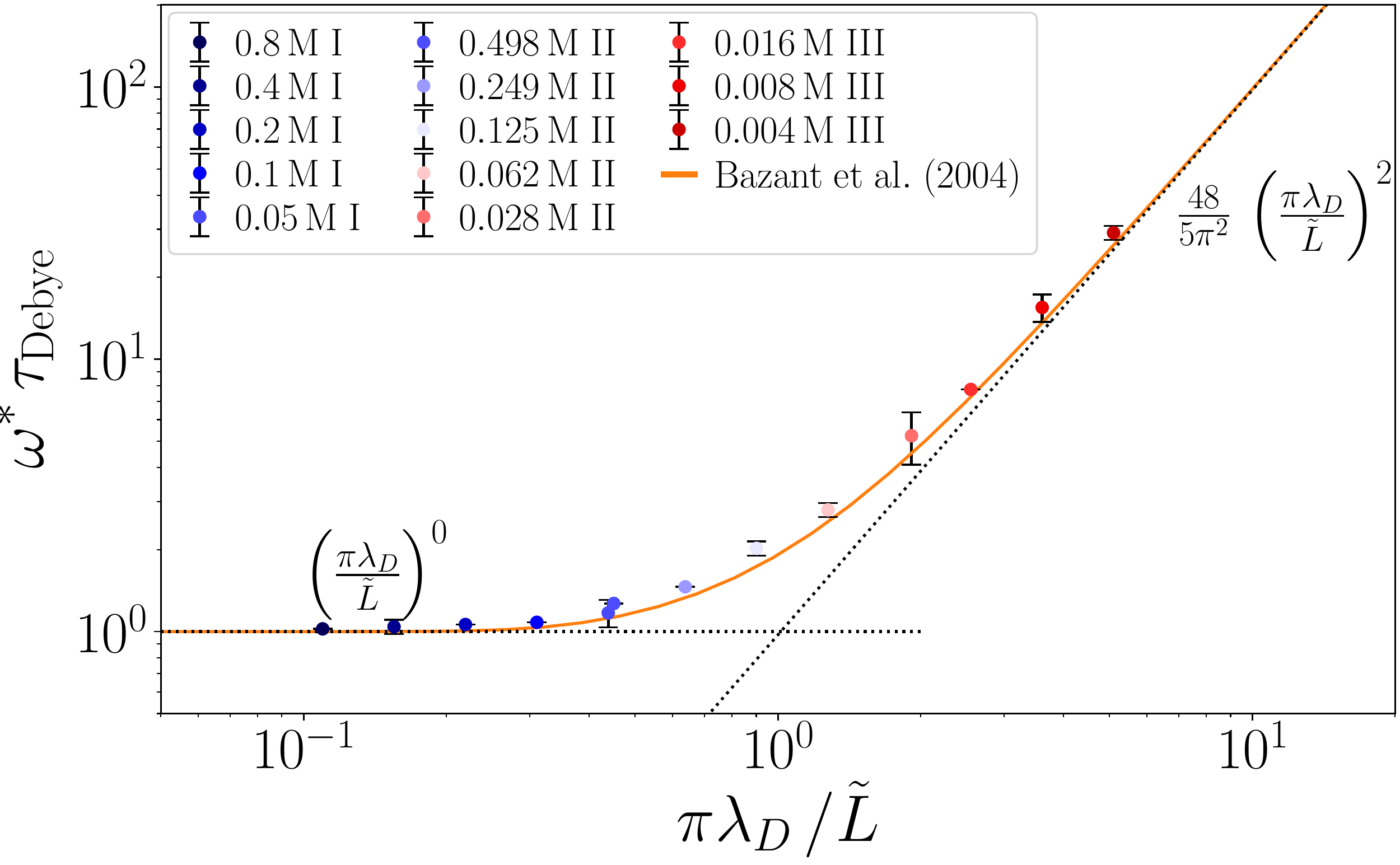}
    \end{center}
    \caption{
    Crossover frequency $\omega^*$, defined by Eq.~\ref{eq:DefOmegaStar}, scaled by the Debye time $\tau_{\rm Debye}$ as a function of the ratio $\pi \lambda / \tilde{L}$ between the relevant length scales for electrostatic screening ($\lambda_D$) and diffusion ($\tilde{L}/\pi$), for a wide range of concentrations $C_s$ and several distances $\tilde{L}$ between the walls (see Table~\ref{tab:systems}). The red solid line corresponds to $\omega^*= 1/\tau_{\rm Bazant}$ using Eq.~\eqref{eq:tauBazant} (from the results of Ref.~\citenum{bazant_diffuse-charge_2004}), while the two dotted lines indicate the scalings expected for small and large values of $\pi \lambda_D / \tilde{L}$.
    \label{fig:crossoverscaling}
    }
\end{figure}

A quantitative analysis of the crossover frequency is provided in Fig.~\ref{fig:crossoverscaling}, which shows $\omega^*$, defined by Eq.~\eqref{eq:DefOmegaStar}, scaled by the Debye time, as a function of the ratio $\pi \lambda_D / \tilde{L}$ between the relevant length scales for electrostatic screening ($\lambda_D$) and diffusion ($\tilde{L}/\pi$), for all the considered systems. In the thin double-layer limit ($\pi \lambda_D / \tilde{L} \ll 1$), $\omega^*\tau_{\rm Debye}\approx 1$. For larger values of the ratio $\pi \lambda_D / L$, the transition depends not only on the bulk relaxation time but also on the confining distance $L$. For $\pi \lambda_D / \tilde{L} \gg 1$, the quadratic scaling of $\omega^*\tau_{\rm Debye}$ with $\pi \lambda_D / \tilde{L} $  shows that the transition between low and high conductivity regimes occurs for $\omega^*\propto1/\tau_{\rm diff}$: In this limit, the Debye length is much larger than the confining distance and the relevant process for charge transport is the diffusion of ions between the walls. The crossover between two regimes occurs for $\pi \lambda_D / \tilde{L} \approx 1$, as expected.

Interestingly, all our numerical results are very well described by an analytical prediction (also shown in Fig.~\ref{fig:crossoverscaling}) following from the work of Bazant \textit{et al.} who solved the Poisson--Nernst--Planck (PNP) equation to analyze the charge dynamics in a capacitor~\cite{bazant_diffuse-charge_2004}. Here, we do not consider the response of an electrolyte confined between metallic electrodes to a voltage step, but rather the response to an applied electric field, which corresponds to boundary conditions of fixed surface charge rather than surface potential. In the linear response regime, to which this section is restricted, the conductivity corresponds to the fluctuations of the current under conditions of vanishing applied field, hence vanishing surface charge. This case can be recovered in Ref.~\citenum{bazant_diffuse-charge_2004} by considering the limit $\delta\to\infty$ in their result Eq.~(50) for the time scale characterizing the total charge in a half-cell. Indeed, while the effective thickness $\lambda_S$ introduced in the electrostatic boundary conditions (Eq.~(14) in Ref.~\citenum{bazant_diffuse-charge_2004}) has a different interpretation in their work, the limit $\lambda_S\to\infty$ amounts to imposing a zero electric field at the surface of the solid. Using the notation of the present work for the distance between the walls ($L$ instead of $2L$), and scaling the characteristic time defined in Eq.~(50) of Ref.~\citenum{bazant_diffuse-charge_2004} by the Debye relaxation time (instead of the $RC$ time), one obtains in this limit:
\begin{equation}
    \frac{\tau_{\rm Bazant}}{\tau_{\rm Debye}} =  1 - \displaystyle \frac{\tilde{L}}{4\lambda_D} \coth{\frac{\tilde{L}}{4\lambda_D}}\sech{\frac{\tilde{L}}{2\lambda_D}}.
    \label{eq:tauBazant}
\end{equation}
The solid line in Fig.~\ref{fig:crossoverscaling} corresponds to the assumption $\omega^*= 1/\tau_{\rm Bazant}$, which is in excellent agreement with the simulation results. One recovers in particular the two limiting regimes, $\tau_{\rm Bazant}\sim\tau_{\rm Debye}$ and $\tau_{\rm Bazant}\sim \frac{48}{5\pi^2}\tau_{\rm diff}\approx 0.97\tau_{\rm diff}$ for small and large values of $\pi \lambda_D / \tilde{L} $, respectively. In the intermediate region, $\tau_{\rm Bazant}$ is close to $\tau_{RC}$, but the corresponding range of $\pi \lambda_D / \tilde{L} $ is rather limited.
The time scale in Eq.~\eqref{eq:tauBazant} corresponds to the dynamics of the total charge of the half-cell, but other time scales could be determined in Ref.~\citenum{bazant_diffuse-charge_2004}, such as for the local charge density of the electrolyte at the surface of the electrodes. The good agreement between our results for the crossover frequency and the former time (instead of the latter or other measures of the charge dynamics) suggests that the evolution of the total charge of a half-pore better reflects the slower fluctuations of the total ionic current.

The shape of $\sigma(\omega)$ depends on the functional form of the current ACF (see Eq.~\eqref{eq:GK_conductivity_BD}): A mono-exponential decay results in a Lorentzian function for the frequency-dependent conductivity, but the current ACF may display more complex features, such as several exponential modes with different weights. Another, somewhat more surprising observation is that the simulation results are well described by the mean-field PNP equation even for concentrations as high as 0.8~M. At such concentrations, this theory is not expected to be accurate, as for example the Debye screening length becomes comparable to the size of ions and water molecules, which questions the validity of continuous descriptions.  Nevertheless, the above discussion further supports the relevance of the present approach to predict the frequency-dependent conductivity from current fluctuations in Brownian dynamics simulation.

%%%%%%%%%%%%%%%%%%%%%%%%%%%%%%%%%%%%%%%%%%
\section{Conclusion}

Using Brownian dynamics simulations, we investigated the effects of confinement, adsorption on surfaces and ion-ion interactions on the response of electrolyte solutions confined between parallel walls to oscillating electric fields. The frequency-dependent conductivity characterizing the linear response always decays from a bulk-like behavior at high frequency to a vanishing conductivity at low frequency due to the confinement of the charge carriers by the walls. We discussed the characteristic features of the crossover between the two regimes, most importantly how the crossover frequency depends on the confining distance and the salt concentration, and the fact that adsorption on the walls may lead to significant changes in both the high- and low-frequency regions. Conversely, our results illustrate the possibility to obtain information on diffusion between the walls, charge relaxation and adsorption on the walls, by analyzing the frequency-dependent conductivity. The interplay between adsorption on the walls and ion-ion interactions, which were considered separately in the present work, will be of particular interest.

We obtained the frequency-dependent conductivity from the equilibrium fluctuations of the electric current, using the Green--Kubo relation appropriate for overdamped dynamics, which differs from the standard one for Newtonian or underdamped Langevin dynamics. This expression (Eq.~\eqref{eq:GK_conductivity_BD}) highlights the contributions of the underlying Brownian fluctuations and of the interactions of the particles between them and with external potentials (here, confining walls). While already known in the literature, the practical use of this relation seems very limited to date, beyond the static limit $\omega\to 0$ to determine the effective diffusion coefficient or the static conductivity. We hope that the present work will motivate readers to consider it to investigate the response of other Brownian systems to various time- or space-dependent external drivings, \textit{e.g.} to determine transport coefficients from equilibrium simulations along the lines developed in Ref.~\citenum{cheng_computing_2020} for heat conductivity, or the frequency-dependent response of soft matter systems to magnetic fields~\cite{kreissl_frequency_2021}. As a natural extension of this case of insulating walls, one might for example consider the dynamics of a Brownian electrolyte between metallic walls (see Ref.~\citenum{cats_capacitance_2021} for static properties), or more generally the effect of dielectric jumps between the confined liquid and the surrounding media. In addition, future studies should also assess the effect of the coupling of ions with hydrodynamic flows, which was not considered in this work. In particular, the conductivity in the directions parallel to the surfaces (not discussed here) may deviate more significantly from the bulk response.

The nonequilibrium response further allows to investigate the transition from linear to nonlinear regimes upon increasing the magnitude of the electric field. Even though we did not explore this regime in the case of interacting particles, previous work has shown the variety of behaviors that can arise by driving Brownian particles in opposite directions~\cite{dzubiella_lane_2002, chakrabarti_reentrance_2004}, from charge correlations and hydrodynamic interactions in confined electrolytes under strong electric fields (see \textit{e.g.} Ref.~\citenum{lobaskin_diffusive-convective_2016}), or when the central region of the liquid becomes depleted from ions due to their migration towards the walls~\cite{bazant_diffuse-charge_2004}. Compared to the Green--Kubo approach for the linear response, which provides the full frequency spectrum from equilibrium fluctuations, the nonequilibrium study requires running a simulation for each frequency. For a more efficient investigation of the nonlinear regime, it might be possible to take advantage of the recent developments based, on large deviation theory, to predict the response far from equilibrium from the fluctuations in nonequilibrium steady-states~\cite{lesnicki_field_2020,lesnicki_molecular_2021}.

%%%%%%%%%%%%%%%%%%%%%%%%%%%%%%%%%%%%%%%%%%%%%%%%%%%%%%%%%%%%%%%%%%%%%%%%%%%%%%%%%%%%
\appendix

%%%%%%%%%%%%%%%%%%%%%%%%%%%%%%%%%%%%%%%%%%
\section{Green--Kubo formula for Brownian dynamics}
\label{sec:Appendix:GK}

\newcommand{\rme}{\mathrm{e}}
\newcommand{\ri}{\mathrm{i}}

We provide in this appendix a derivation of Eq.~\eqref{eq:GK_conductivity_BD}, already obtained by Felderhof and Jones in Ref.~\citenum{felderhof_linear_1983}, using arguments similar to those in Ref.~\citenum{joubaud_langevin_2015}. The basic idea is to write out the perturbation of the probability distribution of the system (described by the positions $\vect{R}=\left\{ \vect{r}_i \right\}_{i=1\dots 2N}$ corresponding to the trajectories of ions following Eq.~\eqref{eq:BD}), which determines the linear response of the electric current introduced in Eq.~\eqref{eq:JelBD}:
\begin{align}
  J_\mathrm{el}({\bf R},t)
  & = \sum_i \beta D_i q_i F_{i,z}({\bf R},t) \nonumber \\
  & = \sum_i \beta D_i q_i F^0_{i,z}({\bf R}) + E_0 \sin(\omega t) \beta \sum_i D_i q_i^2 \nonumber \\
  & = J_\mathrm{el}^0({\bf R}) + E_0 \sin(\omega t) \beta \sum_i D_i q_i^2,
  \label{eq:Jcontributions}
\end{align}
where we denote by~$F^0$ the forces for the system in the equilibrium situation~$E_0 = 0$, and introduced the corresponding current $J_\mathrm{el}^0$.

For a given strength of the external field~$E_0$, the stationary probability distribution $\rho_{E_0}({\bf R},t)$ corresponds to a periodic cycle which has the same period as the external forcing. In particular, $\rho_{E_0}({\bf R},t) = \rho_{E_0}({\bf R},0)$. The space-time probability distribution satisfies the following Fokker--Planck equation with periodic boundary conditions in time:
\begin{equation}
    \left[-\partial_t \, + \mathcal{L}_{\rm eq}^\dagger + E_0 \sin(\omega t) \mathcal{L}_{\rm pert}^\dagger\right] \rho_E({\bf R},t) = 0,
    \label{eq:FP}
\end{equation}
where
\begin{equation}
\mathcal{L}_{\rm eq} = \sum_i D_i (\beta F^0_{i,z}({\bf R}) \partial_{z_i} + \Delta_{{\bf r}_i}),
\quad
\mathcal{L}_{\rm pert} = \beta \sum_i D_i q_i \partial_{z_i},
\end{equation}
are respectively the generator of the equilibrium dynamics and the generator of the external perturbation, with adjoints
\begin{equation}
  \mathcal{L}_{\rm eq}^\dagger f = \sum_i D_i \left(\Delta_{{\bf r}_i} f - \beta \partial_{z_i} \left[ F^0_{i,z} f\right] \right),
  \quad
  \mathcal{L}_{\rm pert}^\dagger = -\mathcal{L}_{\rm pert}.
    \label{eq:LFPE}
\end{equation}
In the absence of external field, the equilibrium distribution $\rho_0({\bf R})$ does not depend on time and satisfies Eq.~\eqref{eq:FP} for $E_0=0$. We write $\rho_0({\bf R}) = \mathrm{e}^{-\beta \mathcal{U}({\bf R})}$, where~$\mathcal{U}$ is the potential energy function of the system (shifted in order to incorporate the normalization that the probability measure should sum to~1), so that~${\bf F}_i^0 = -\nabla_{\bf r_i} \mathcal{U}$. 

Linear response theory corresponds to identifying the leading order term~$\nu_1({\bf R},t)$ in the expansion of~$\rho_{E_0}({\bf R},t)$ in powers of~$E_0$:
\begin{equation}
\rho_{E_0}({\bf R},t) = \rho_0({\bf R}) + E_0 \nu_1({\bf R},t) + E_0^2 \nu_2({\bf R},t) + \mathrm{O}(E_0^3) \; .
\label{eq:rhocontributions}
\end{equation}
Upon identifying terms with the same powers of~$E_0$ in Eq.~\eqref{eq:FP}, the first non-trivial condition reads
\begin{equation}
\left(-\partial_t \, + \mathcal{L}_{\rm eq}^\dagger \right) \nu_1({\bf R},t) = - \sin(\omega t) \mathcal{L}_{\rm pert}^\dagger \rho_0({\bf R}).
\end{equation}
Given that~$\nu_1$ is real valued, and that the time dependence of the right hand side of the previous equation involves only~$\rme^{\pm \ri \omega t}$, and in fact~$\mathrm{Im}(\rme^{\ri \omega t})$, we look for a solution of the form
\begin{equation}
\nu_1({\bf R},t) = \mathrm{Im}\left(\widetilde{\nu}_1({\bf R}) \, \rme^{\ri \omega t}\right).
\end{equation}
The function~$\widetilde{\nu}_1$ satisfies
\begin{equation}
\left(-\ri \omega + \mathcal{L}_{\rm eq}^\dagger \right)\widetilde{\nu}_1({\bf R}) = -\mathcal{L}_{\rm pert}^\dagger \rho_0({\bf R}).
\end{equation}
A simple computation shows that the right hand side is equal to~$\beta J_{\rm el}^0 \rho_0$. Therefore,
\begin{equation}
\widetilde{\nu}_1 = -\beta \left(\ri \omega - \mathcal{L}_{\rm eq}^\dagger \right)^{-1} \left(J_{\rm el}^0  \rho_0\right)
\; .
\label{eq:TildeNuOne}
\end{equation}
The function~$\widetilde{\nu}_1$ is well defined as it can be shown, similarly to what is done in Ref.~\citenum{joubaud_langevin_2015}, that the operator~$\ri \omega - \mathcal{L}_{\rm eq}^\dagger$ can be inverted on spaces of functions with average~0 (which is the case here for the function~$J_{\rm el}^0 \rho_0$), while the adjoint operator~$\ri \omega - \mathcal{L}_{\rm eq}$ can be inverted on spaces of functions with average~0 with respect to~$\rho_0$.

The linear response of a real valued observable~$\theta$ with average~0 with respect to~$\rho_0$ is 
\begin{equation}
  \int \theta({\bf R}) \nu_1({\bf R},t) \, d{\bf R} = \mathrm{Im}\left( s_\theta(\omega) \rme^{\ri \omega t}\right) \; , 
  \label{eq:sthetadef}
\end{equation}
where, using~\eqref{eq:TildeNuOne} and performing an integration by part in order to let the operator act on~$\theta$,
\begin{align}
  s_\theta(\omega)
  & = \int \theta({\bf R}) \widetilde{\nu}_1({\bf R}) \, d{\bf R} 
  \nonumber \\
  & = -\beta \int \theta({\bf R}) \left[\left(\ri \omega - \mathcal{L}_{\rm eq}^\dagger\right)^{-1} \left( J_{\rm el}^0 \, \rho_0\right)\right]\!\!({\bf R}) \, d{\bf R} \nonumber \\
  & = -\beta \int \left[\left(\ri \omega - \mathcal{L}_{\rm eq}\right)^{-1}\theta\right]\!\!({\bf R}) \, J_{\rm el}^0({\bf R}) \, \rho_0({\bf R}) \, d{\bf R}.
\end{align}
Since $\rme^{\tau \mathcal{L}_{\rm eq}} \to 0$ as $\tau \to +\infty$ on spaces of functions with average~0 with respect to~$\rho_0$, it holds
\begin{align}
& \left(\ri \omega - \mathcal{L}_{\rm eq}\right) \int_0^{+\infty} \rme^{-\ri \omega \tau} \rme^{\tau \mathcal{L}_{\rm eq}} \, d\tau \nonumber \\
& \qquad \qquad = -\int_0^{+\infty} \frac{d}{d\tau} \left[ \rme^{-\ri \omega \tau} \rme^{\tau \mathcal{L}_{\rm eq}} \right] d\tau = \mathrm{Id}.
\end{align}
This leads to the following operator identity on functions with average~0 with respect to~$\rho_0$:
\begin{equation}
\label{eq:inverse_as_time_integral}
\left(\ri \omega - \mathcal{L}_{\rm eq}\right)^{-1} = \int_0^{+\infty} \rme^{-\ri \omega \tau} \rme^{\tau \mathcal{L}_{\rm eq}} \, d\tau.
\end{equation}
Therefore,
\begin{align}
  s_\theta(\omega) & = -\beta \int \int_0^{+\infty} \rme^{-\ri \omega \tau} \left(\rme^{\tau \mathcal{L}_{\rm eq}}\theta\right)\!\!({\bf R}) \, J_{\rm el}^0({\bf R}) \, \rho_0({\bf R}) \, d{\bf R} \, d\tau 
  \nonumber \\
  & = -\beta \int_0^{+\infty} \rme^{-\ri \omega \tau} %C_{\theta,J_{\rm el}^0}(\tau) 
  \left\langle \theta(\tau) J_{\rm el}^0(0) \right\rangle_{0}
  \, d\tau,
  \label{eq:stheta}
\end{align}
where
\begin{align}
\left\langle \theta(\tau) J_{\rm el}^0(0) \right\rangle_{0} 
  &= %C_{\theta,J_{\rm el}^0}(\tau) & = 
  \int \left(\rme^{\tau \mathcal{L}_{\rm eq}}\theta\right)\!\!({\bf R}) \, J_{\rm el}^0({\bf R}) \, \rho_0({\bf R}) \, d{\bf R} %\nonumber \\
  %& = \left\langle \theta(\tau) J_{\rm el}^0(0) \right\rangle_{0}
  \label{eq:ctheta}
\end{align}
is the correlation function between $\theta$ and $J_{\rm el}^0$ obtained by averaging over trajectories of the equilibrium dynamics started from initial conditions distributed according to~$\rho_0$.

Using the decompositions Eqs.~\eqref{eq:Jcontributions} and~\eqref{eq:rhocontributions}, as well as the normalization $\int \rho_{0}({\bf R}) \, d{\bf R}=1$, we can now express the stationary current to linear order in $E_0$ as
\begin{align}
  J_{\rm el}(t) &= \int J_{\rm el}({\bf R},t) \rho_{E_0}({\bf R},t) \, d{\bf R}
  \nonumber \\
  & = \int J_{\rm el}^0({\bf R}) \rho_{0}({\bf R}) \, d{\bf R}
  + E_0 \left[ \sin(\omega t) \beta \sum_i D_i q_i^2 \right.
  \nonumber \\
    & \left. \qquad \qquad + \int J_{\rm el}^0({\bf R},t) \nu_1({\bf R},t) \, d{\bf R}\right] + \mathrm{O}(E_0^2) \; .
\end{align}
The first term vanishes, while the second one can be rewritten using Eqs.~\eqref{eq:sthetadef} to~\eqref{eq:ctheta} for $\theta=J_{\rm el}^0$. One finally obtains (see Eq.~\eqref{eq:LinearResponse})
\begin{equation}
  \lim_{E_0 \to 0} \frac{J_{\rm el}(t)}{E_0} = \mathrm{Im}\left( \rme^{\ri \omega t} V \tilde{\sigma}(\omega) \right)  
\end{equation}
with an explicit expression of the complex conductivity 
\begin{equation}
  \tilde{\sigma}(\omega) = \frac{\beta}{V} \sum_i D_i q_i^2 - \frac{\beta}{V}  \int_0^{+\infty} \rme^{-\ri \omega \tau} \left\langle J_{\rm el}^0(\tau) J_{\rm el}^0(0) \right\rangle_{0}\, d\tau
  \; ,
\end{equation}
which completes the proof of Eq.~\eqref{eq:GK_conductivity_BD}.

%%%%%%%%%%%%%%%%%%%%%%%%%%%%%%%%%%%%%%%%%%
\section{Spectral resolution of the Fokker--Planck equation for ideal particles}
\label{sec:Appendix:FP}

Here, we solve the FP Eq.~\eqref{eq:FP} in the case of ideal particles in the absence of applied electric field ($E=0$), in order to compute the expectation value of the spectral density of the current appearing in the Green--Kubo formula Eq.~\eqref{eq:GK_conductivity_BD}. For this simple case of ions only experiencing the force due to the confining walls, the Fokker--Planck operator Eq.~\eqref{eq:LFPE} is a sum of independent 1-particle operators $\mathcal{L}_\mathrm{eq}^0 = \sum_i L_{\mathrm{eq},i}^0$ with $L_{\mathrm{eq},i}^0$ acting only on functions of~${\bf r}_i$. Moreover, the $2N-$particle density factorizes as a product of 1-particle densities. In the geometry we consider, it is in fact sufficient to consider the 1-particle density $\rho(z,t)$ as a function of $z$ and $t$ only, and the 1-particle current in the direction $z$ perpendicular to the interface, with $J_\mathrm{el}^0 = \beta D q F_z = - \beta D q U'(z)$ (see Eq.~\eqref{eq:Walls} for the definition of~$U$). In this context, and in view of the derivation leading to~\eqref{eq:stheta}, the spectral density of this current can be expressed as
\begin{equation}
    \int_0^{\infty} \left\langle  J_\mathrm{el}^0(0) J_\mathrm{el}^0(t) \right\rangle_0 \mathrm{e}^{-\mathrm{i} \omega t}  \, \diff t \ = \ \int_0^{\infty} \left\langle J_\mathrm{el}^0, \mathrm{e}^{t L_{\mathrm{eq}}^0}J_\mathrm{el}^0 \right\rangle_0 \mathrm{e}^{-\mathrm{i} \omega t}  \, \diff t,
\end{equation}
where the scalar product is
\begin{equation}
\langle f,g \rangle_0 = \int_{-L/2}^{L/2} f(z) g(z) \rho_0(z) \, dz,
\end{equation}
and the generator of the evolution acts on functions of~$z$ only:
\begin{equation}
L_\mathrm{eq}^0 f = D \left( \partial_z^2 f- \beta (\partial_z U) \partial_z f\right).
\end{equation}
This operator is endowed with non-flux boundary conditions $\partial_z f(\pm L/2) = 0$ at the walls. 

The generator is negative and symmetric for the scalar product~$\langle \cdot,\cdot\rangle_0$, and can therefore be decomposed as  
\begin{equation}
    L_\mathrm{eq}^0 = -\sum_\alpha \left| \phi_{\alpha} \right\rangle \lambda_\alpha \left\langle \phi_{\alpha}  \right|,
\end{equation}
with $\lambda_\alpha \geq 0$ and $\phi_{\alpha}$ the eigenvalues and eigenvectors of $L_\mathrm{eq}^0$, the eigenfunctions being normalized as~$\langle \phi_\alpha,\phi_\alpha\rangle_0 = 1$. Note that the eigenfunction associated with the smallest eigenvalue in absolute value, namely~0, is a constant function. The other eigenvalues are negative. Since~$\langle J_\mathrm{el}^0, \mathbf{1} \rangle_0 = 0$, we obtain, in view of an operator identity similar to~\eqref{eq:inverse_as_time_integral},
\begin{equation}
    \int_0^{\infty} \left\langle  J_\mathrm{el}(0) J_\mathrm{el}(t) \right\rangle_0 \mathrm{e}^{-\mathrm{i} \omega t}  \, \diff t \ = \ \displaystyle \sum_\alpha \frac{\left|  \left\langle J_\mathrm{el} | \phi_{\alpha} \right\rangle_0 \right|^2}{-\lambda_\alpha + i \omega}.
    \label{eq:PSDfromEigen}
\end{equation}

The direct numerical diagonalization of~$L_\mathrm{eq}^0$ is prone to instabilities because of divergences at the boundaries that are incompatible with Neumann conditions. We circumvent this problem by a standard transformation relying on a change of unknown function. We introduce to this end the shifted potential $\mathfrak{U}(z)$ such that $\rho_0(z) = \mathrm{e}^{-\beta \mathfrak{U}(z)}$, and write
\begin{equation}
    \phi_\alpha(z) = \mathrm{e}^{\beta \mathfrak{U}(z)/2 } \psi_\alpha(z)
    \; .
\end{equation}
The functions~$\psi_\alpha$ are then eigenfunctions of the stationary Schr\"odinger equation
\begin{equation}
    H \psi_\alpha = \lambda_\alpha \psi_\alpha, \qquad \int \psi_\alpha^2 = 1,
\end{equation}
for a Hamiltonian with an effective potential:
\begin{equation}
H = D \, \left( -\partial_z^2 + \beta \, W \right), 
\quad
W(z) = \frac{\beta}{4} \left( \mathfrak{U}'(z) \right)^2 - \frac{1}{2} \mathfrak{U}''(z).
\end{equation}
Neumann boundary conditions on the eigenfunctions~$\phi_\alpha$ translate, for regular potentials, into Robin boundary conditions for $\psi_\alpha$, namely $\partial_z \psi_\alpha(\pm L/2) + \psi_\alpha(\pm L/2) \beta \mathfrak{U}'(\pm L/2)/2 = 0$. For singular potentials at the boundaries, these conditions simplify to Dirichlet boundary conditions $\psi_\alpha(\pm L/2) = 0$.

Numerically, in order to avoid instabilities in the dynamics, we consider the problem on the restricted interval $\left[-L/2+0.7\sigma_{\rm w},L/2-0.7\sigma_{\rm w}\right]$, divided using $N+2$ equally  spaced points $z^i$ with lattice spacing $h = (L-1.4\sigma_{\rm w})/(N+1)$. The values of the eigenfunctions are set to~0 at the end points of the interval, and only the values at the~$N$ interior points are sought. In this setting, the operator~$H$ is represented by the following $N\times N$ tridiagonal matrix obtained by a central finite difference scheme for the one-dimensional Laplacian operator:
\begin{equation}
H^{i,j} = -D\frac{\delta^{i+1,j} - 2 \delta^{i,j} + \delta^{i-1,j} }{h^2} + \beta DW(z^i) \delta^{i,j},
\end{equation}
with $\delta^{k,l}=1$ if $k=l$ and 0 otherwise. This matrix is symmetric and we diagonalize it numerically using a standard NumPy linear algebra library. The output is a finite set $\left\lbrace \lambda_{\alpha}, \psi_{\alpha}^i\right\rbrace_{0 \leq \alpha \leq N-1}$ with ordered real eigenvalues and corresponding normalized orthonormal functions, which we use to approximate Eq.~\eqref{eq:PSDfromEigen}. In practice, we use $N=3000$ which does not differ from the result for $N=2000$ or 2500 by more than 1\% of the conductivity over the whole frequency range considered.

%%%%%%%%%%%%%%%%%%%%%%%%%%%%%%%%%%%%%%%%%%
\section{Nonlinear regime}
\label{sec:Appendix:NL}

\begin{figure}[ht!]
   \begin{center}
       \includegraphics[width=0.45\textwidth]{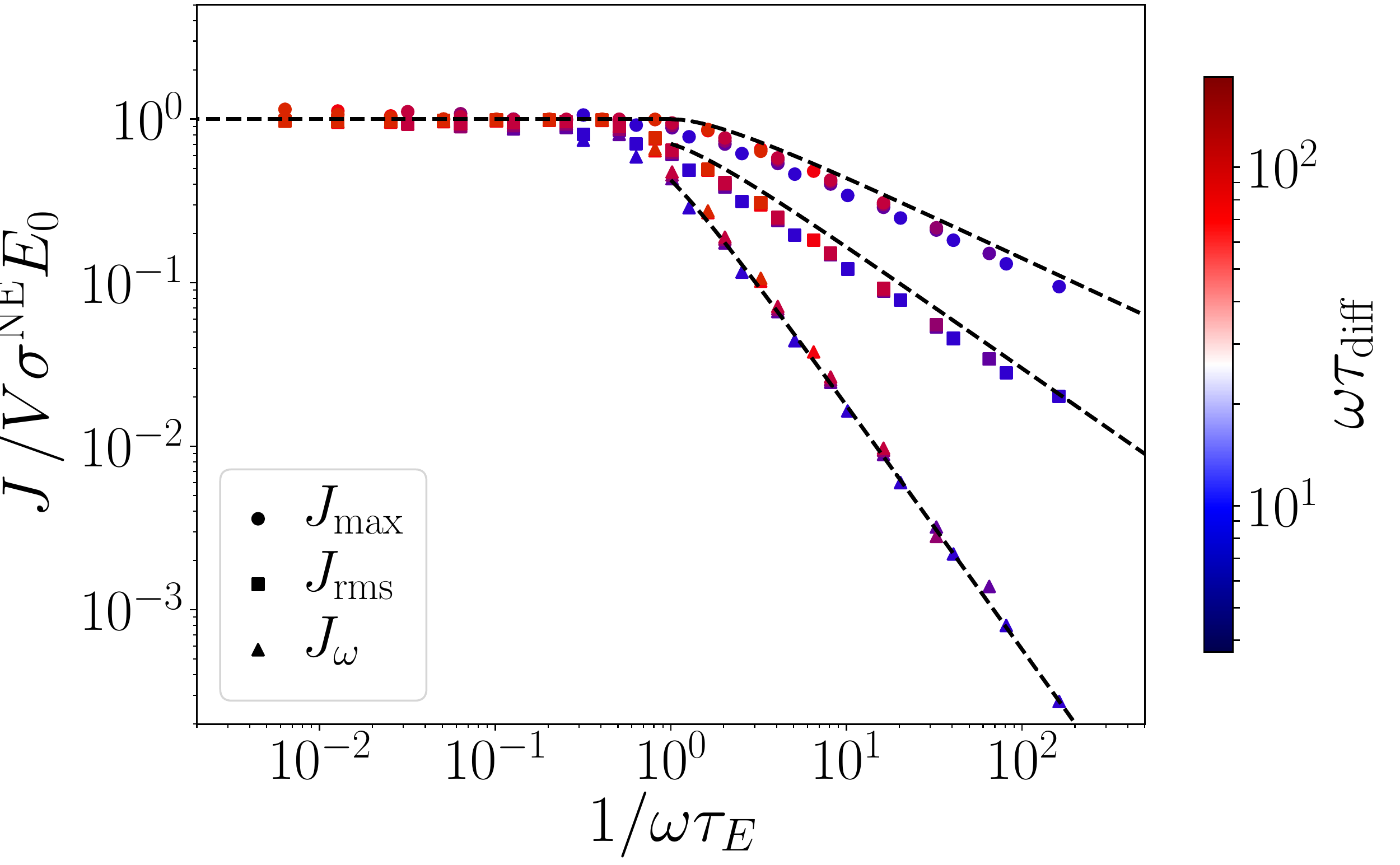} 
   \end{center}
     \caption{
    Maximum value of the nonequilibrium steadystate current, $J_{\rm max}$ (see Eq.~\eqref{eq:Jmax}), root-mean-square current, $J_{\rm rms}$ (see Eq.~\eqref{eq:Jrms}) and Fourier component of the current at the frequency of the applied field, $J_\omega$ (defined as $J_\mathrm{el}(\omega)$ in Eq.~\eqref{eq:Jomega}), as a function of the inverse of the frequency scaled by the migration time $\tau_{E}$ (see Eq.~\eqref{eq:tauE}). The three currents are normalized by the maximal current for ideal particles in the absence of confinement by the walls (bulk case), $\sigma^{\mathrm{NE}}E_0$, as in Fig.~\ref{fig:NEQ_currents}, for all the simulations corresponding to Fig.~\ref{fig:SigmaEandOmega}; only the frequency limited to the range $\omega\tau_{\rm diff}\gtrsim 2$ are shown. The lines are the predictions Eqs.~\eqref{eq:Jmaxmodel}, \eqref{eq:Jrmsmodel} and~\eqref{eq:Jomegamodel} of the simple model presented in this Appendix.
    }
     \label{fig:nonlinearJ}
\end{figure}

In Section~\ref{sec:Results:EandOmega} we examined the stationary current $J_\mathrm{el}(t)$ in the presence of an oscillating external field $E(t)=E_0\sin(\omega t)$ and the conditions under which this response is nonlinear in the perturbation. In particular, we noted that for sufficiently large fields and low frequencies the current reaches a maximum $J_{\rm max}$ at a finite time $t_{\rm max}<T/4$, with $T=2\pi/\omega$ the period of the applied field, before vanishing before $T/2$; a symmetric observation can be made for negative currents and fields for the second half of the period. Here we propose a simple model to analyze this nonlinear response in more detail, which we quantify by the maximum current
\begin{equation}
    J_{\max} = \max_{0 \leq t \leq T} \left\lbrace \, J_\mathrm{el}(t) \, \right\rbrace
    \label{eq:Jmax}
\end{equation}
and the root-mean-square current,
\begin{equation}
    J_{\mathrm{rms}}
        \ = \
        \sqrt{ 2 \, \left\langle \, J_\mathrm{el}^2(t) \, \right\rangle }
        \ = \ 
        \left( \frac{2}{T}\int_0^{T} J_\mathrm{el}^2(t) \, \diff t \right)^{1/2}
    \label{eq:Jrms}
\end{equation}
which do not only involve the response at the frequency of the applied field, in addition to the Fourier component of the current at the frequency of the applied field, $J_\omega$ (defined as $J_\mathrm{el}(\omega)$ in Eq.~\eqref{eq:Jomega}). The simulation results for these quantities are shown in Fig.~\ref{fig:nonlinearJ}, as a function of the inverse of the frequency scaled by the migration time $\tau_{E}$. For $\omega\tau_E \gg 1$, all measures of the current coincide with the current for ideal particles in the absence of confinement by the walls, while for $\omega\tau_E < 1$ they differ (sufficiently large fields and/or sufficiently low frequencies), suggesting that the nonlinear features are related to the fact that the ions reach the walls within the period of the field.

In order to simplify the discussion, we further assume that cations and anions have the same diffusion coefficient $D$. For large applied fields, the motion of the ions is dominated by the migration of the ions and we can neglect the effect of diffusion. For sufficiently low frequencies such that $\tau_E < T/4 = \pi / 2\omega$ (see Eq.~\eqref{eq:tauE}), the ions reach the walls before the applied field takes its maximum value and then do not move until the field is reversed at $T/2$. At the beginning of each period, the ions start moving from $\pm L/2$ (depending on the sign of their charge). Solving their equation of motion $\dot{z}_\pm(t) = \pm \beta D e E_0 \sin(\omega t)$, we find that the time required to reach the opposite wall is
\begin{equation}
    t_{\rm max} = \frac{\arccos \left( 1 - \omega \tau_E \right)}{\omega} 
    \label{eq:t_max_appC}
\end{equation}
and that the current is maximum at that time, with
\begin{equation}
    J_{\rm max} = V \sigma^\mathrm{NE} E_0 \sin(\omega  t_{\rm max})
    \; .
\end{equation}
In this high field limit, the current should then vanish abruptly until the field is reversed. Fig.~\ref{fig:NEQ_currents} shows that this is not the case even for the largest applied field considered here. The continuous decay to zero after $t_{\rm max}$ comes from the particles that arrive later due to the dispersion of the ions arising from the thermal fluctuations (which also explains why the current is also reversed slightly before the reversal of the field). This second phase therefore depends not only on $\tau_E$ but also on the diffusive time scale $\tau_{\rm diff}$. In order to make analytical predictions, we do not take this into account and simply model the rise and decay of the current as:
\begin{equation}
    J_\mathrm{el}(t) =
    \left\lbrace
    \begin{matrix}
    \displaystyle J_{\max}  \sin\left( \frac{\pi }{2 t_{\max}} t \right), &
    t \in \left[0, \,  2 t_{\max}\right], \\
    0,  \ & t \in \left[2 t_{\max} , \, T/2\right], 
    \end{matrix}
    \right.
\end{equation}
and an opposite current for the second half-period. While this model is a rather crude approximation, it allows us to obtain analytical predictions for $J_{\rm max}$, $J_{\rm rms}$ and $J_\omega$. Specifically, we find that they can be expressed as 
\begin{align}
    J_{\rm max}  & = V \sigma^{\mathrm{NE}} E_0 \ \sqrt{1 - \left( 1 - \omega\tau_E\right)^2},
    \label{eq:Jmaxmodel} \\
    J_{\rm rms}  & = J_{\max} \, \displaystyle  \sqrt{ \frac{2}{\pi} \arccos \left(1 - \omega \tau_E \right) }, 
    \label{eq:Jrmsmodel} \\
    J_\omega &=  J_{\max} \displaystyle \frac{ 2 \, \sqrt{\left( 1 - \omega \tau_E \right)^2-\left( 1 - \omega \tau_E \right)^4} \, \arccos\left( 1 - \omega \tau_E \right) }{ \left(\pi/2\right)^2 - \arccos^2\left( 1 - \omega \tau_E \right) }.
    \label{eq:Jomegamodel}
\end{align}

These predictions are also indicated in Fig.~\ref{fig:nonlinearJ}. The good agreement with the simulation results for the three measures of the nonlinear behavior confirms that the fact that the ions reach the wall within the period is the main origin of the latter.

%%%%%%%%%%%%%%%%%%%%%%%%%%%%%%%%%%%%%%%%%%
\section*{Acknowledgements}

The authors thank Marie Jardat, Emmanuel Trizac, Ivan Palaia and Roland Netz for useful discussions. This project received funding from the European Research Council under the European Union’s Horizon 2020 research and innovation program (project SENSES, grant Agreement No. 863473 and project EMC2, grant agreement No. 810367) and from the Agence Nationale de la Recherche (project SINEQ, grant ANR-21-CE40-0006). This work was performed with the support of the Institut des Sciences du Calcul et des Données (ISCD) of Sorbonne University (IDEX SUPER 11-IDEX-0004).

%%%%%%%%%%%%%%%%%%%%%%%%%%%%%%%%%%%%%%%%%%
\section*{Author declarations}

\subsection*{Conflict of interest}

The authors have no conflicts to disclose.

\subsection*{Author contributions}

\textbf{Th\^e Hoang Ngoc Minh:} Conceptualization (equal); Formal Analysis (equal); Investigation (lead); Methodology (equal); Writing/Original Draft Preparation (equal); Validation (equal); Writing/Review \& Editing (supporting);
\textbf{Gabriel Stoltz:} Conceptualization (equal); Formal Analysis (equal); Funding Acquisition (supporting); Methodology (equal); Supervision (supporting); Validation (equal); Writing/Review \& Editing (supporting);
\textbf{Benjamin Rotenberg:} Conceptualization (lead); Formal Analysis (equal); Funding Acquisition (lead); Investigation (supporting); Methodology (equal); Supervision (lead); Validation (equal); Writing/Original Draft Preparation (equal); Writing/Review \& Editing (lead).

%%%%%%%%%%%%%%%%%%%%%%%%%%%%%%%%%%%%%%%%%%
\section*{Data availability}

The data that support the findings of this study are available from the corresponding author upon reasonable request.

%%%%%%%%%%%%%%%%%%%%%%%%%%%%%%%%%%%%%%%%%%%%%%%%%%%%%%%%%%%%%%%%%%%%%%%%%%%%%%%%%%%%
\bibliographystyle{aipnum4-1}
%\bibliography{biblio}


\end{document}